\documentclass[aps,prd,twocolumn,showpacs,nofootinbib,superscriptaddress,10pt,floatfix]{revtex4-2}

\usepackage{amsmath,amssymb}
\usepackage{graphicx}
\usepackage{physics}
\usepackage{slashed}
\usepackage{color}
\usepackage[dvipsnames]{xcolor}
\usepackage{booktabs}
\usepackage{array}

\usepackage{hyperref}
\hypersetup{%
	colorlinks = true,
	linkcolor  = MidnightBlue,
	urlcolor   = BrickRed,
	citecolor  = MidnightBlue
}%

\begin{document}
	
	\title{Self-stimulated avalanche in superradiant axion clouds}
	
	\author{Zhen-Hong Lyu}
	\email{lyuzhenhong@itp.ac.cn}
	\affiliation{Institute of Theoretical Physics, Chinese Academy of Sciences (CAS), Beijing 100190, China}
	\affiliation{School of Physical Sciences, University of Chinese Academy of Sciences (UCAS), Beijing 100049, China}
	
	\author{Rong-Gen Cai}
	\email{caironggen@nbu.edu.cn}
	\affiliation{Institute of Fundamental Physics and Quantum Technology, \& School of Physical Science and Technology, Ningbo University, Ningbo, 315211, China}
	
	\author{Jing Liu}
	\email{liujing@ucas.ac.cn}
	\affiliation{International Centre for Theoretical Physics Asia-Pacific, University of Chinese Academy of Sciences, Beijing 100190, China}
	\affiliation{Taiji Laboratory for Gravitational Wave Universe (Beijing/Hangzhou), University of Chinese Academy of Sciences, Beijing 100049, China}
	
	\begin{abstract}
        Boson clouds formed via superradiance around spinning black holes offer a novel probe of ultralight particles. We show that such gravitational atoms can undergo a self-stimulated avalanche: a coherent quadrupolar transition is seeded by external gravitational waves and amplified by self-generated radiation feedback. We formulate an effective two-level description that captures the logistic population transfer and the resulting delayed gravitational-wave pulse with a characteristic envelope. This cooperative emission mechanism opens a new observational avenue into the ultralight dark sector.
    \end{abstract}
	
\maketitle
	
{\section{Introduction}} 
Ultralight bosons emerge as well-motivated extensions of the Standard Model and compelling dark matter candidates~\cite{PhysRevLett.40.223,Arvanitaki:2009fg,Marsh:2015xka,Hui:2016ltb,Ferreira:2020fam}, yet their weak nongravitational couplings render them difficult to detect. Black holes (BHs) provide a purely gravitational probe: superradiance can extract angular momentum from a rotating BH, forming a long-lived, macroscopically occupied cloud~\cite{Zeldovich:1971ffh,Starobinskii:1973vzb,Press:1972zz,Brito:2015oca,Arvanitaki:2010sy,Arvanitaki:2014wva,Baumann:2019eav}. This process is governed by the gravitational coupling $\alpha\equiv GM\mu$ (in natural units $\hbar=c=1$), where $M$ is the BH mass and $\mu$ the boson mass, and is efficient for $\alpha\sim\mathcal{O}(0.01-0.1)$. Characterised by a hydrogenic structure, this system is referred to as a gravitational atom and exhibits rich phenomenology through distinctive gravitational-wave (GW) and electromagnetic signatures~\cite{Brito:2017wnc,Ghosh:2018gaw,LIGOScientific:2021rnv,Yuan:2022bem,Guo:2022mpr,Liu:2024mzw,Kang:2024trj,Guo:2025pea,Takahashi:2026ruw,Guo:2026yiz,Bian:2025gwcosmology,Rosa:2017ury,Ikeda:2018nhb,Spieksma:2023vwl,Chen:2019fsq,Chen:2021lvo,Chen:2022kzv}, serving as a powerful probe for ultralight bosons~\cite{Stott:2018opm,Fernandez:2019qbj,Davoudiasl:2019nlo,Ng:2020ruv,Liu:2021llm,Saha:2022hcd,Cheng:2022jsw,Duque:2023seg,Witte:2024drg,Hoof:2024quk,Caputo:2025oap,Aswathi:2025nxa,Bai:2025yxm,Tomaselli:2025zdo,Roy:2025qaa,Lyu:2025lue,Lyu:2025nsd,Li:2026dls}. Notably, a recent analysis of the LIGO--Virgo--KAGRA catalog reported tentative evidence for such boson clouds when informed by superradiance priors~\cite{Roy:2025qaa}.

In particular, the discrete spectrum and large occupation number of the cloud allow transitions between atomic levels to occur as coherent, macroscopic processes. Much of the recent literature has focused on transitions driven by the companion's tidal field in a binary system. In analogy with atomic physics, this perturbation can resonantly mix bound levels via Landau-Zener transitions, ionize the cloud, and backreact on the orbit~\cite{Baumann:2018vus,Baumann:2019ztm,Baumann:2021fkf,Baumann:2022pkl,Berti:2019wnn,Zhang:2018kib,Zhang:2019eid,Boskovic:2024fga,Takahashi:2021yhy,Takahashi:2024fyq,Cao:2023fyv,Tong:2022bbl,Tomaselli:2023ysb,Tomaselli:2024bdd,Tomaselli:2024dbw,Guo:2024iye,Guo:2025ckp,Peng:2025zca,Li:2025qyu,Kyriazis:2025fis,Tomaselli:2025jfo,Kim:2025wwj,Liang:2026njr,Li:2026wds}. The resulting characteristic features in the GW waveform have been proposed as signatures of ultralight bosons in BH binaries~\cite{Baumann:2022pkl,Tomaselli:2024dbw,DellaMonica:2025zby,Dyson:2025dlj,Li:2025ffh,Xu:2026cky,Xu:2026aic}.

Beyond companion-induced tidal transitions, incident GWs introduce a qualitatively distinct perturbation through a self-stimulated mechanism. The process is initiated when a weak external wave seeds a coherent interference quadrupole, which in turn radiates stimulated GWs that couple back into the same transition, locking the atom--wave system into an autonomous feedback loop. Thus, instead of merely driving negligible Rabi oscillation, this tiny perturbation unlocks a cooperative avalanche in which a minority component in the lower level amplifies the transition of the remaining cloud. This unfolds as the gravitational analogue of superfluorescence~\cite{PhysRev.93.99,PhysRevA.11.1507,PhysRevA.14.1169,GROSS1982301}. Just as an inverted atomic ensemble develops a macroscopic polarization that feeds back to release stored energy in a short electromagnetic flash, the cloud's interference quadrupole self-amplifies to rapidly deplete the upper level, ultimately emitting a characteristic GW pulse~\footnote{Conventionally, superfluorescence is initiated by spontaneous quantum fluctuations in an inverted ensemble. Here the analogy refers to the subsequent cooperative emission stage: a weak external GW prepares the initial coherence, after which the self-generated radiation drives the collective avalanche.}.

Stimulated gravitational emission from superradiant axion clouds has been explored in several complementary settings~\cite{Dupuis:2018dhs,Liu:2024mzw,An:2026ugw}.
Ref.~\cite{Dupuis:2018dhs} studied stimulated graviton production along orbiting null trajectories near the BH. More recently, Ref.~\cite{An:2026ugw} showed that a superradiant cloud can amplify an ambient stochastic GW background through stimulated transitions between cloud levels. 
Ref.~\cite{Liu:2024mzw} revealed the self-stimulated mechanism and derived the avalanche dynamics through a flux-based estimate of the self-generated field. This estimate, however, treated the self-generated field using the transverse-traceless (TT) interaction, which neglects the constrained metric components present inside the cloud that cannot be eliminated by imposing the TT gauge. Here we formulate the interaction in terms of the gauge-invariant tidal tensor, equivalently the complete stress--metric vertex with a conserved transition source, and determine the feedback from the full retarded response.
We develop an effective two-level description of a boson cloud coupled to its gravitational radiation. We show that the radiative self-field drives collective population transfer, while the conservative response shifts the coherence phase. After the seed has passed, and when lower-level absorption is negligible, the resulting dynamics yields a logistic population avalanche and a delayed, quasi-monochromatic GW pulse with a hyperbolic-secant envelope set by the collective transition timescale. We will use a metric with “mostly plus” signature $(-, +, +, +)$ and work in natural units with $\hbar = c = 1$.

\vspace{1.8\baselineskip}
	
\section{{Coupled atom--wave system}}
Consider a real scalar field $\phi$ of mass $\mu$ around a BH of mass $M$. Its nonrelativistic cloud is controlled by the gravitational coupling $\alpha$. Using the ansatz $\phi=(2\mu)^{-1/2}\Psi e^{-i\mu t}+{\rm c.c.}$, the Klein–Gordon equation in the Kerr background reduces, at leading order in $\alpha$, to the effective Schrödinger equation $i\partial_t\Psi=[-\nabla^2/(2\mu)-\alpha/r]\Psi$. The bound states are labelled by $\ket{n\ell m}$, with wave functions $\Psi_{n\ell m}=R_{n\ell}(r)Y_{\ell m}(\theta,\varphi)e^{-i(\omega^\Psi-\mu) t}$ and a ``Bohr radius'' $r_B=(\mu\alpha)^{-1}$. The corresponding complex eigenfrequencies can be written as $\omega^\Psi_{n\ell m}=\omega_{n\ell m}+i\gamma_{n\ell m}$, where the energy spectrum reads~\cite{Baumann:2019eav}
\begin{equation}
\omega_{n\ell m}=\mu\left[1-\frac{\alpha^2}{2n^2}-F_{n\ell}\alpha^4+h_{n\ell}\tilde{a}m\alpha^5+\mathcal{O}(\alpha^6)\right]\,.
\label{eq:main_ga_spectrum}
\end{equation}
Here $\tilde a=J/(GM^2)$ is the dimensionless BH spin, with $J$ the BH angular momentum. The imaginary part $\gamma_{n\ell m}\propto\mu\alpha^{4\ell+5}$ is the superradiant growth or absorption rate of the level. The detailed expression of $\gamma_{n\ell m}$ and the coefficients $F_{n\ell}$ and $h_{n\ell}$ are given in Appendix~\ref{sec:sm_gravitational_atom}.


Focusing on a pair of relevant levels, we denote the superradiantly populated upper state by $\ket{1}$ and the lower state into which the cloud transitions by $\ket{2}$. The two-level cloud is then described by
\begin{equation}
    \Psi = \sqrt{N}\,\bigl[c_1(t)\psi_1 e^{-i\omega_1 t} + c_2(t)\psi_2 e^{-i\omega_2 t}\bigr]\,,
\end{equation}
with $|c_1|^2+|c_2|^2=1$ and an initially inverted population, $c_1(0)\simeq 1$ and $c_2(0)\simeq 0$. Here $N$ is the occupation number and $M_c = \mu N$ is the cloud mass, and $\omega_{a}=\omega_{n_a\ell_am_a}-\mu$ is the real binding energy of level $a$ ($a=1,2$).

To linear order, GWs couple to the scalar field through a weak dynamical perturbation of the metric, $g_{\mu\nu}=\bar{g}_{\mu\nu}+h_{\mu\nu}$, where the unperturbed background $\bar g_{\mu\nu}$ encodes the stationary BH field that determines the bound states, while $h_{\mu\nu}$ denotes the additional field sourced by the externally incident wave and the cloud itself. We adopt the comoving Fermi coordinates $(t,\bf{x})$, whose origin is located at the centre-of-mass of the unperturbed BH-cloud. In the nonrelativistic limit, the tidal interaction potential up to quadrupole order is written as~\cite{Baumann:2018vus}
\begin{equation}
    V_{\rm tidal}(t,\mathbf{x})=-\frac{\mu}{2}h^{00}\simeq \frac{\mu}{2}E_{ij}(t)x_ix_j\,,
\end{equation}
where
\begin{align}
E_{ij}(t)&=R_{0i0j}(t,\mathbf{0})\nonumber\\
&=\frac{1}{2}\left.\left(\partial_i\dot h_{0j}+\partial_j\dot h_{0i}-\partial_i\partial_j h_{00}-\ddot h_{ij}\right)\right|_{\bf x = 0}\label{eq:full_tidal_tensor}
\end{align}
is the linearized tidal tensor. The two-level equation reads~\footnote{We omit the diagonal matrix elements $V_{aa}$ below since constant diagonal shifts are absorbed into the level frequencies, whereas their oscillating components are nonresonant and are discarded at leading rotating-wave order.}
\begin{equation}
i\dot c_1=V_{12}(t)e^{i\omega_0t}c_2\,,\quad
i\dot c_2=V_{21}(t)e^{-i\omega_0t}c_1\,.\label{eq:two_level_equation}
\end{equation}
Here $V_{ab}(t)=\mu E_{ij}(t)\mathcal Q_{ij}^{ab}/2$ is the tidal matrix element, $\mathcal Q_{ij}^{ab}\equiv\langle a|x_ix_j|b\rangle\equiv\int\mathrm{d}^3\mathbf x\,\psi_a^*(\mathbf x)x_ix_j\psi_b(\mathbf x)$, $\omega_0\equiv \omega_1 - \omega_2>0$ is the energy splitting of the two levels. The indices $a,b=1,2$ label the two bound states. The complete stress--metric coupling and its tidal reduction are derived in Appendix~\ref{sec:sm_two_level}.

The tidal field contains an external seed and a self-generated contribution,
\begin{equation}
E_{ij}(t)=E_{ij}^{\mathrm{seed}}(t)+E_{ij}^{\mathrm{self}}(t)\,.
\end{equation}
We model the near-resonant incident GW as $h_{ij}^{\mathrm{seed}}=h_E(t)e_{ij}^{(\lambda)} e^{i\mathbf{k}\cdot\mathbf{x}-i\omega_{\mathrm s}t}+\mathrm{c.c.}$, where $\omega_{\mathrm s}\simeq \omega_0$, $|\mathbf{k}|=\omega_{\mathrm s}$, and $e_{ij}^{(\lambda)}$ is the TT polarization tensor. For an envelope $h_E(t)$ that varies slowly compared with the wave period, the seeded tidal field is
\begin{equation}
    E_{ij}^{\mathrm{seed}} = -\frac{1}{2}\ddot{h}_{ij}^{\mathrm{seed}} \simeq \frac{\omega_{\mathrm s}^2}{2} h_E(t)e_{ij}^{(\lambda)} e^{-i\omega_{\mathrm s}t}+\mathrm{c.c.}\,.\label{eq:seed_tidal_response}
\end{equation}
In contrast, the self-generated tidal response $E_{ij}^{\mathrm{self}}(t)$ is evaluated in the presence of the cloud source and generally contains scalar, vector, and tensor contributions. Applying the scalar-vector-tensor (SVT) decomposition to the metric perturbation gives
\begin{equation}
    E_{ij}^{\mathrm{self}}(t)=\left.-\frac12\ddot h_{ij}^{\mathrm{self},\mathrm{TT}}+\partial_i\partial_j\Phi-\frac12\delta_{ij}\ddot\Theta+\partial_{(i}\dot\Xi_{j)}\right|_{\bf x=0}\,.\label{eq:tidal_svt_decomposition}
\end{equation}
Here $\Phi$ and $\Theta$ are gauge-invariant scalar potentials, $\Xi_i$ is a gauge-invariant transverse vector, and $h_{ij}^{\mathrm{self},\mathrm{TT}}$ is the transverse-traceless metric perturbation. Their definitions and field equations are given in Appendix~\ref{sec:sm_svt_response}. At the leading weak-field order considered here, the dynamical response is governed by the linearized Einstein equations. The TT component obeys
\begin{equation}
    (-\partial_t^2+\nabla^2) h_{ij}^{\mathrm{self},\mathrm{TT}}=-16\pi G\Pi_{ij}^{\mathrm{TT}}\,,
    \label{eq:tt_wave}
\end{equation}
where $\Pi_{ij}^{\mathrm{TT}}$ is the TT projection of the spatial components of the conserved effective stress tensor. The scalar and vector potentials are determined by Poisson constraints sourced by its density, momentum, and spatial trace. Solving these constraints together with the retarded TT equation determines the self-field that enters the tidal coupling.

The incident tidal field Eq.~\eqref{eq:seed_tidal_response} separates into positive- and negative-frequency components,
\begin{equation}
E_{ij}^{\mathrm{seed}} = E_{ij}^{\mathrm{seed},(+)}(t)+E_{ij}^{\mathrm{seed},(-)}(t)\,,
\end{equation}
where $E_{ij}^{\mathrm{seed},(+)}(t)= \omega_{\mathrm s}^2 h_E(t)e_{ij}^{(\lambda)} e^{-i\omega_{\mathrm s}t}/2$ and $E_{ij}^{\mathrm{seed},(-)}(t)=[E_{ij}^{\mathrm{seed},(+)}(t)]^*$. The self-field admits an analogous decomposition. The interference between the two cloud levels produces stress-tensor components oscillating at the transition frequencies $\pm\omega_0$.
Since both the metric response and the tidal tensor are linear in the source at the order retained, the two frequency components can therefore be treated independently, giving the resonant self-field
\begin{equation}
E_{ij}^{\mathrm{self}}(t)=E_{ij}^{\mathrm{self},(+)}(t)+E_{ij}^{\mathrm{self},(-)}(t)\,.
\end{equation}

The matrix elements inherit the frequency decomposition of the tidal tensors. In Eq.~\eqref{eq:two_level_equation}, the rotating-wave approximation retains $V_{12}^{(+)}$ in the equation for $c_1$ and $V_{21}^{(-)}=[V_{12}^{(+)}]^*$ in the equation for $c_2$. They split into seed and self parts,
\begin{equation}
V_{12}^{(+)}(t)=V_{12}^{(+),\mathrm{seed}}(t)+V_{12}^{(+),\mathrm{self}}(t)\,,
\end{equation}
with seed contribution
\begin{equation}
V_{12}^{(+),\mathrm{seed}}(t)\equiv
\frac12\Omega_{R}(t)e^{-i\omega_{\mathrm s}t}\,.
\end{equation}
Here $\Omega_{R}(t)\equiv\mu\omega_{\mathrm s}^2Q_{ij}^{12}e_{ij}^{(\lambda)}h_E(t)/2$ is the complex seed Rabi amplitude. The symmetric trace-free (STF) quadrupole $Q_{ij}^{ab}\equiv\mathcal Q_{ij}^{ab}-\delta_{ij}\mathcal Q_{kk}^{ab}/3$ appears because the incident tide is traceless.

With the slowly varying cloud amplitudes $c_{1,2}(t)$ held fixed during the field response, the positive-frequency interference source is proportional to $\zeta(t)e^{-i\omega_0t}$, where $\zeta\equiv c_2^*c_1$ is the two-level coherence. The self-field inherits this dependence by linearity, allowing us to write (see Appendices~\ref{sec:sm_svt_response} and~\ref{sec:sm_self_energy})
\begin{equation}
V_{12}^{(+),\mathrm{self}}(t)\equiv\frac12\Sigma_{12}\zeta(t)e^{-i\omega_0t}\,.
\end{equation}
The coefficient $\Sigma_{12}$ is the resonant gravitational self-energy of the transition.
With detuning $\delta\equiv\omega_{\mathrm s}-\omega_0$, the combined seed and self-field contributions give
\begin{align}
i\dot c_1&=\frac12\left[\Omega_{R}(t)e^{-i\delta t}+\Sigma_{12}c_1c_2^*\right]c_2\,,\\
i\dot c_2&=\frac12\left[\Omega_{R}^*(t)e^{i\delta t}+\Sigma_{12}^*c_1^*c_2\right]c_1\,.\label{eq:main_effective_two_level_0}
\end{align}
These equations close the feedback between the cloud and its gravitational field: the incident wave prepares a two-level coherence, the coherence sources a metric perturbation, and the resulting self-field acts back on the same transition. The next section evaluates $\Sigma_{12}$ and identifies the TT on-shell contribution that drives population transfer.

\section{{Self-stimulated dynamics}}

\subsection{Self-energy and feedback dynamics}

Since the self-field is the sum of its constrained and TT parts, the self-energy coefficient splits as
\begin{equation}
    \Sigma_{12}=\Sigma_{\mathrm C,12}+\Sigma_{\mathrm{TT},12}\,.
\end{equation}
The constraint contribution $\Sigma_{\mathrm C,12}$ is sourced by the scalar and vector potentials $(\Phi,\Theta,\Xi_i)$ of the SVT decomposition \eqref{eq:tidal_svt_decomposition}. These fields satisfy Poisson-type constraint equations, whose solutions do not describe propagating radiation, so the constraint self-energy is \textit{conservative},
\begin{equation}
\operatorname{Im}\Sigma_{\mathrm C,12}=0\,.
\end{equation}
Eq.~\eqref{eq:main_effective_two_level_0} therefore shows that the constraint sector does not contribute to the dissipative dynamics; the explicit proof is given in Appendix~\ref{sec:sm_self_energy}.

The TT response, by contrast, contains both a conservative and a \textit{dissipative} part. At leading quadrupole order, the TT contribution follows from the retarded solution of the wave equation~\eqref{eq:tt_wave} in Fourier space, which carries the propagation pole $\mathbf k^2-\omega_0^2$ (see Appendix~\ref{sec:sm_self_energy}),
\begin{equation}
\Sigma_{\mathrm{TT},12}
\simeq
-4\pi G\mu M_c\omega_0^4
\int\frac{\mathrm{d}^3\mathbf k}{(2\pi)^3}
\frac{
Q_{ij}^{12}
\Lambda_{ij,kl}(\hat{\mathbf k})
Q_{kl}^{21}
}
{\mathbf{k}^2-(\omega_0+i0)^2}\,.
\end{equation}
Here $\mathbf k$ is the spatial wave vector of the Fourier mode, $\hat{\mathbf k}\equiv\mathbf k/|\mathbf k|$ is its unit direction, and $\Lambda_{ij,kl}(\hat{\mathbf k})=\mathcal P_{i(k}\mathcal P_{l)j}-\frac12\mathcal P_{ij}\mathcal P_{kl}$ is the TT projector with $\mathcal P_{ij}=\delta_{ij}-\hat k_i\hat k_j$. The retarded prescription separates the response into a principal-value part and an on-shell contribution:
\begin{equation}
\frac{1}{\mathbf{k}^2-(\omega_0+i0)^2}
=
\operatorname{PV}
\frac{1}{\mathbf{k}^2-\omega_0^2}
+i\pi\delta(\mathbf{k}^2-\omega_0^2)\,.
\end{equation}
The imaginary part samples the radiative shell $|\mathbf k|=\omega_0$, and its integral gives
\begin{equation}
\Gamma_{\mathrm{eff}}\equiv-\operatorname{Im}\Sigma_{\mathrm{TT},12}=\frac{2G}{5}\mu M_c\omega_0^5Q_{ij}^{12}Q_{ij}^{21}\,.
\label{eq:geff_definition}
\end{equation}
Here we used that $\operatorname{Im}\Sigma_{\mathrm C,12}=0$.
Eq.~\eqref{eq:geff_definition} is the leading radiative contribution obtained in the quadrupole limit. The conservative contribution to $\Sigma_{12}$, consisting of the constrained response and the TT principal-value part, is instead evaluated using the complete stress--metric coupling in Appendix~\ref{sec:sm_self_energy}, since it is sensitive to the finite spatial structure of the source.

Combining the constraint and TT contributions, we parameterise the resonant self-energy as
\begin{equation}
\Sigma_{12}\equiv-\Delta-i\Gamma_{\mathrm{eff}}\,.
\end{equation}
The effective resonant equations reduce to
\begin{align}
    i\dot c_1 &=\frac{1}{2}\left[\Omega_R e^{-i\delta t}
    -(\Delta+i\Gamma_{\rm eff})c_1c_2^*\right]c_2\,,\\
    i\dot c_2 &=\frac{1}{2}\left[\Omega_R^* e^{+i\delta t}
    -(\Delta-i\Gamma_{\rm eff})c_1^*c_2\right]c_1\,,
    \label{eq:main_effective_two_level}
\end{align}
Here $\Delta=-\left(\Sigma_{\mathrm C,12}+\operatorname{Re}\Sigma_{\mathrm{TT},12}\right)$ is the conservative level shift, which combines the constrained-sector contribution and the TT principal-value part; it shifts the phase of the atomic coherence. The imaginary term $\Gamma_{\rm eff}$ is in quadrature ($\pi/2$ out of phase) with the atomic coherence and therefore defines the self-stimulated growth.

\begin{figure}[t]
    \centering
    \includegraphics[width=.9\columnwidth]{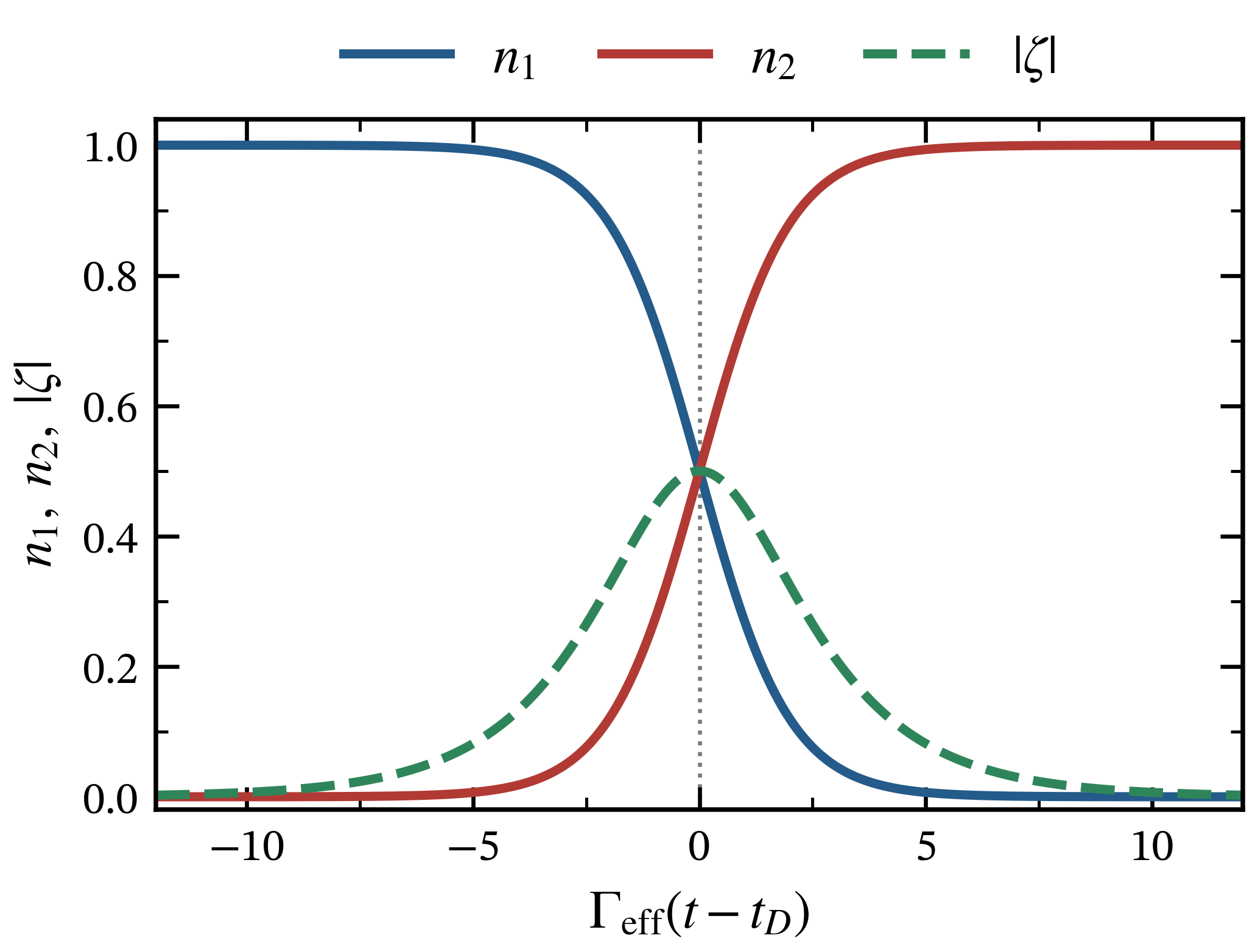}
    \caption{Universal self-stimulated dynamics in the dimensionless time $\Gamma_{\rm eff}(t-t_D)$, centred at the half-population time $t_D$. The upper-level population $n_1$ and lower-level population $n_2$ follow the logistic profiles of Eq.~\eqref{eq:main_logistic} (solid lines), while the coherence magnitude $|\zeta|=|c_1c_2^*|=\tfrac12\operatorname{sech}[\Gamma_{\rm eff}(t-t_D)/2]$ peaks at $t_D$ (dashed line). The curves depend only on Eq.~\eqref{eq:main_logistic} and are independent of the specific transition channel.}
    \label{fig:logistic_avalanche}
\end{figure}

On resonance, once a small lower-level component has been established by the seed, Eq.~\eqref{eq:main_effective_two_level} yields $\dot n_2 = \Gamma_{\rm eff}(1-n_2)n_2$, giving the logistic solution
\begin{equation}
n_2(t)=\frac12\left[1+\tanh\frac{t-t_D}{2t_p}\right]\,,\quad n_1(t)=1-n_2(t)\,,
\label{eq:main_logistic}
\end{equation}
where $n_{a}=|c_{a}|^2$ ($a=1,2$) denotes the level population, $t_p = \Gamma_{\rm eff}^{-1}$, and $t_D$ is defined by $n_2(t_D)=1/2$. 
Consequently, the atomic coherence that sources the tensor field evolves as $|\zeta|=|c_1c_2^*|=\frac12\operatorname{sech}[(t-t_D)/(2t_p)]$, indicating that the stimulated emission peaks when the upper and lower states are equally populated. The logistic solutions are shown in Fig.~\ref{fig:logistic_avalanche}.

The external GW does not modify the universal avalanche dynamics; it only prepares the initial seed. A finite incident wave with Rabi envelope $\Omega_R(t)$ acting from $t=t_{\rm i}$ to $t=t_{\rm s}$, where $t_{\rm i}$ and $t_{\rm s}$ denote the beginning and end of the seed, produces a tiny seed population $n_2^{\rm seed}\simeq \frac14\bigl|\int_{t_{\rm i}}^{t_{\rm s}} \Omega_R(t)e^{-i(\delta-\Delta/2)t}\,\mathrm{d}t\bigr|^2$, which determines the ignition delay $t_D\simeq-t_p\ln{n_2^{\rm seed}}$. The ignition efficiency is sensitive to both the seed detuning $\delta=\omega_{\mathrm s}-\omega_0$ and the self-induced detuning $\Delta/2$ caused by the conservative near-zone field. These combined detuning effects affect only the onset time by altering initial lower-level population, but the subsequent avalanche remains logistic and is governed solely by $\Gamma_{\rm eff}$, as detailed in Appendix~\ref{sec:sm_full_dynamics}.

\subsection{Parametric scaling of the transition rate}

We define the cloud mass fraction $\sigma\equiv M_c/M$ and write the trace-free quadrupole as $Q_{ij}^{12}=r_B^2\,\widehat Q_{ij}^{12}$, where $\widehat Q_{ij}^{12}$ is a dimensionless, channel-dependent coefficient. Writing the level splitting as $\omega_0=\widehat\omega_0\,\mu\alpha^{p}$, where $\widehat\omega_0$ is a dimensionless, level-dependent coefficient and $p=2,4,5$ for Bohr, fine, and hyperfine splittings respectively, the transition rate \eqref{eq:geff_definition} becomes
\begin{equation}
    \Gamma_{\mathrm{eff}}
    =\frac{2}{5}\widehat\omega_0^5\,\|\widehat Q^{12}\|^2\,\sigma\mu\alpha^{5p-3}\,,
    \label{eq:rate_scaling}
\end{equation}
with $\|\widehat Q^{12}\|^2\equiv\widehat Q_{ij}^{12}\widehat Q_{ij}^{21}$. For a Bohr transition ($p=2$) this gives $\Gamma_{\rm eff}\propto\sigma\mu\alpha^7$, whereas fine ($p=4$) and hyperfine ($p=5$) transitions scale as $\propto\sigma\mu\alpha^{17}$ and $\propto\sigma\mu\alpha^{22}$. The fifth power $\omega_0^5$ therefore suppresses fine and hyperfine rates by many orders of magnitude, and we henceforth restrict attention to Bohr transitions.

\subsection{Quadrupole selection rules}

A transition is driven only if its STF quadrupole is nonzero. In the spherical basis the STF matrix element reads $Q_{2q}^{21}\equiv\langle n_2\ell_2 m_2|\,r^2 C_{2q}\,|n_1\ell_1 m_1\rangle$ with $C_{2q}\equiv\sqrt{4\pi/5}\,Y_{2q}$, where $Y_{2q}(\theta,\varphi)$ are the $\ell=2$ spherical harmonics. By the Wigner--Eckart theorem,
\begin{equation}
    Q_{2q}^{21}\propto\mathcal R_{21}
    \begin{pmatrix}\ell_2&2&\ell_1\\0&0&0\end{pmatrix}
    \begin{pmatrix}\ell_2&2&\ell_1\\-m_2&q&m_1\end{pmatrix}\,,
    \label{eq:q21_wigner}
\end{equation}
where $\mathcal R_{21}\equiv\int_0^\infty\mathrm{d}r\,r^4 R_{n_2\ell_2}^*R_{n_1\ell_1}$ is the radial overlap. The $3j$ symbols vanish unless
 \begin{equation}
    |\ell_1-\ell_2|\le 2\le\ell_1+\ell_2\,,\quad
    \ell_1+\ell_2\ \text{even}\,,\quad
    |m_1-m_2|\le 2\,,
    \label{eq:selection_rules}
\end{equation}
i.e.\ $\Delta\ell=0,\pm2$ (same parity) and $\Delta m=0,\pm1,\pm2$, with the further requirement that $\mathcal R_{21}\neq0$ so that the overlap does not vanish accidentally.

\subsection{Lower-level absorption and representative channels}

The self-stimulated dynamics described above applies to an idealized two-level system prepared in an inverted state, $n_1\simeq1$ and $n_2\simeq0$. Superradiance provides exactly such a preparation, leaving the upper level macroscopically occupied. Unlike optical superfluorescence, however, a gravitational atom is not isolated: the BH continuously drains the lower level whenever its single-mode frequency carries a negative imaginary part, $\gamma_{n_2\ell_2 m_2}<0$, which competes with the stimulated transfer. The preceding mechanism is therefore the general feedback framework for an ideal inverted two-level system, while a realistic superradiant cloud additionally requires incorporating the lower-level absorption, which we turn to now.

\begin{figure}[t]
    \centering
    \includegraphics[width=.9\columnwidth]
    {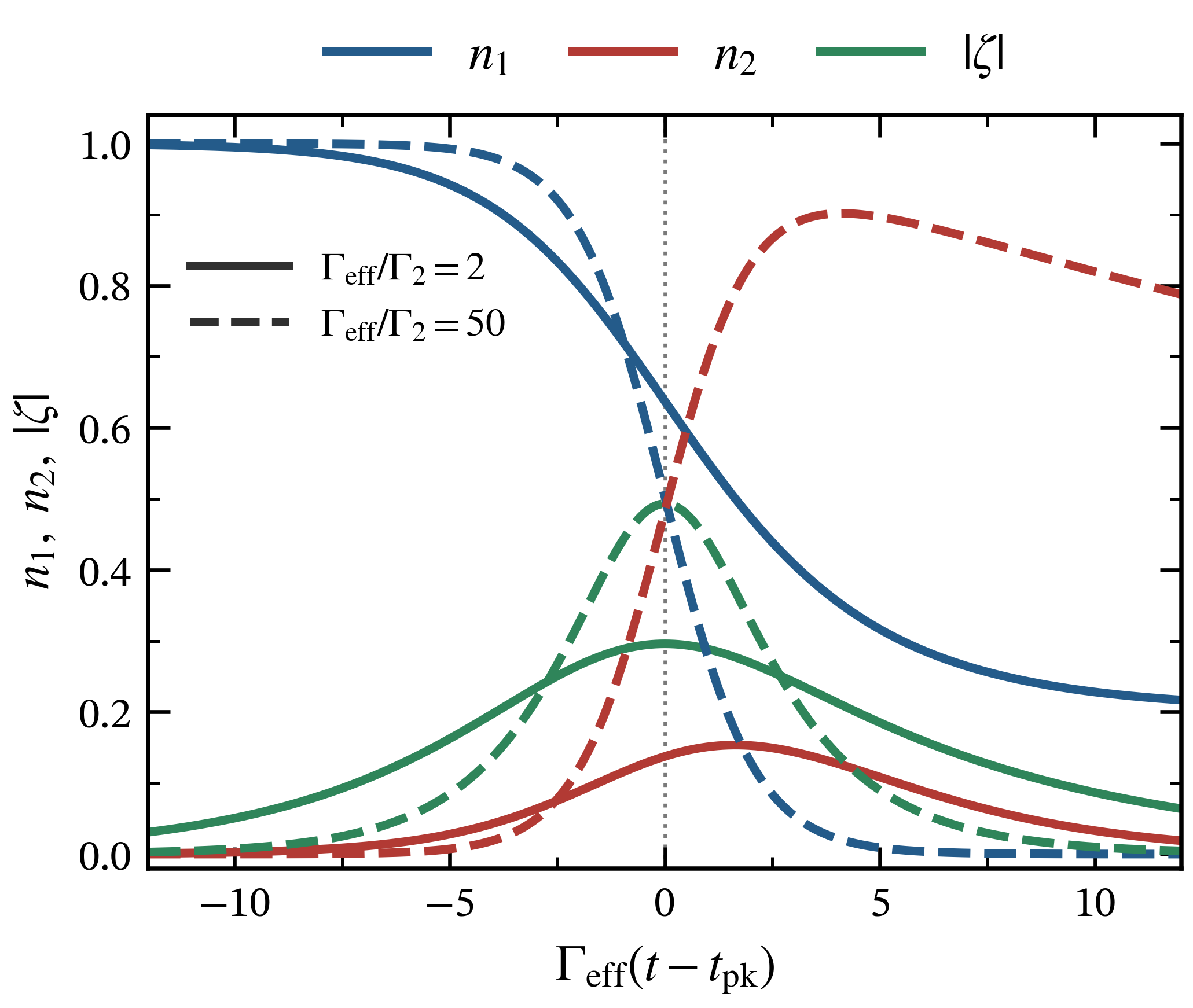}
    \caption{Population dynamics including lower-level absorption.
    Solid and dashed curves correspond to
    $\Gamma_{\rm eff}/\Gamma_2=2$ and $50$,
    respectively. Each evolution is centred at its own $t_{\rm pk}$, defined as the time at which $|\zeta|^2=n_1n_2$ is maximal; the vertical dotted line marks $t=t_{\rm pk}$ for both curves.
    For $\Gamma_{\rm eff}/\Gamma_2=50$, the transition approaches the lossless
    logistic limit, whereas stronger absorption suppresses both the
    transferred population and the peak coherence.}
    \label{fig:loss_modified_dynamics}
\end{figure}

Including this lower-level loss, the population equations become
\begin{equation}
    \dot n_1=-\Gamma_{\mathrm{eff}}n_1n_2\,,\qquad
    \dot n_2=\Gamma_{\mathrm{eff}}n_1n_2-\Gamma_2 n_2\,,
    \label{eq:populations_loss}
\end{equation}
where $\Gamma_2\equiv 2|\gamma_{n_2\ell_2 m_2}|$ is the occupation-number loss rate, with $\gamma_{n\ell m}$ given in Appendix~\ref{sec:sm_gravitational_atom}, scaling as $\Gamma_2\propto\alpha^{4\ell_2+5}$. For the fully inverted initial state $n_1\simeq1$, net growth of the lower level requires
\begin{equation}
    \Gamma_{\mathrm{eff}}>\Gamma_2\,.
    \label{eq:growth_condition}
\end{equation}
The ratio $\Gamma_{\rm eff}/\Gamma_2$ controls not only the onset of the instability but also the efficiency of the population transfer. To compare the pulse profiles independently of their loss-dependent ignition delays, we define $t_{\rm pk}$ for each solution as the time at which $|\zeta|^2=n_1n_2$ reaches its maximum and centre the time coordinate on this peak. We numerically integrate Eq.~\eqref{eq:populations_loss} for two representative loss ratios, with the results shown in Fig.~\ref{fig:loss_modified_dynamics}.

For $\Gamma_{\rm eff}/\Gamma_2=50$, the peak coherence nearly reaches the lossless value $1/2$, and the upper level is almost completely depleted during the burst. In contrast, for $\Gamma_{\rm eff}/\Gamma_2=2$, the peak coherence is reduced to around $0.3$, and a substantial upper-level population survives. Eq.~\eqref{eq:growth_condition} is therefore sufficient for the initial exponential growth, but nearly complete collective transfer requires $\Gamma_{\rm eff}/\Gamma_2\gg1$.

\begin{table}[htbp]
    \caption{Bohr channels on the superradiant ladder. For each dominant upper state, the allowed lower levels and the outcome of the absorption criterion are listed.}
    \label{tab:channel_coefficients}
    \centering
    \scriptsize
    \renewcommand{\arraystretch}{1.8}
    \begin{tabular}{c c c}
        \toprule
        Upper state & Allowed lower levels & Outcome \\
        \midrule
        $\ket{211}$ & --- & $211\to100$ has $\Delta\ell=-1$, forbidden \\
        $\ket{322}$ & $\ket{100},\ \ket{200}$ & quenched by absorption \\
        $\ket{433}$ & $\ket{211},\ \ket{311}$ & quenched by absorption \\
        $\ket{544}$ & $\ket{322},\ \ket{422}$ & viable \\
        \bottomrule
    \end{tabular}
\end{table}

Restricting to Bohr transitions on the superradiant ladder, the dominant upper states are $\ket{211}$, $\ket{322}$, $\ket{433}$, and $\ket{544}$, i.e.\ the $\ell=m$ modes that dominate successive stages of cloud growth. The selection rules \eqref{eq:selection_rules} and the absorption criterion \eqref{eq:growth_condition} together determine the channels, summarised in Table~\ref{tab:channel_coefficients}. The $\ket{211}$ state admits no allowed Bohr channel, since its only lower candidate $211\to100$ has $\Delta\ell=-1$ and is forbidden. The $\ket{322}$ and $\ket{433}$ states decay only to $\ell=0$ and $\ell=1$ lower levels, whose absorption $\Gamma_2\propto\alpha^{4\ell+5}$~\cite{Baumann:2019eav} quenches the avalanche. Only the $\ket{544}$ state decays to weakly absorbed $\ell=2$ lower levels, leaving viable channels.

\section{{GW pulse and sensitivity estimates}}

\begin{figure}[htbp]
    \centering
    \includegraphics[width=.9\columnwidth]{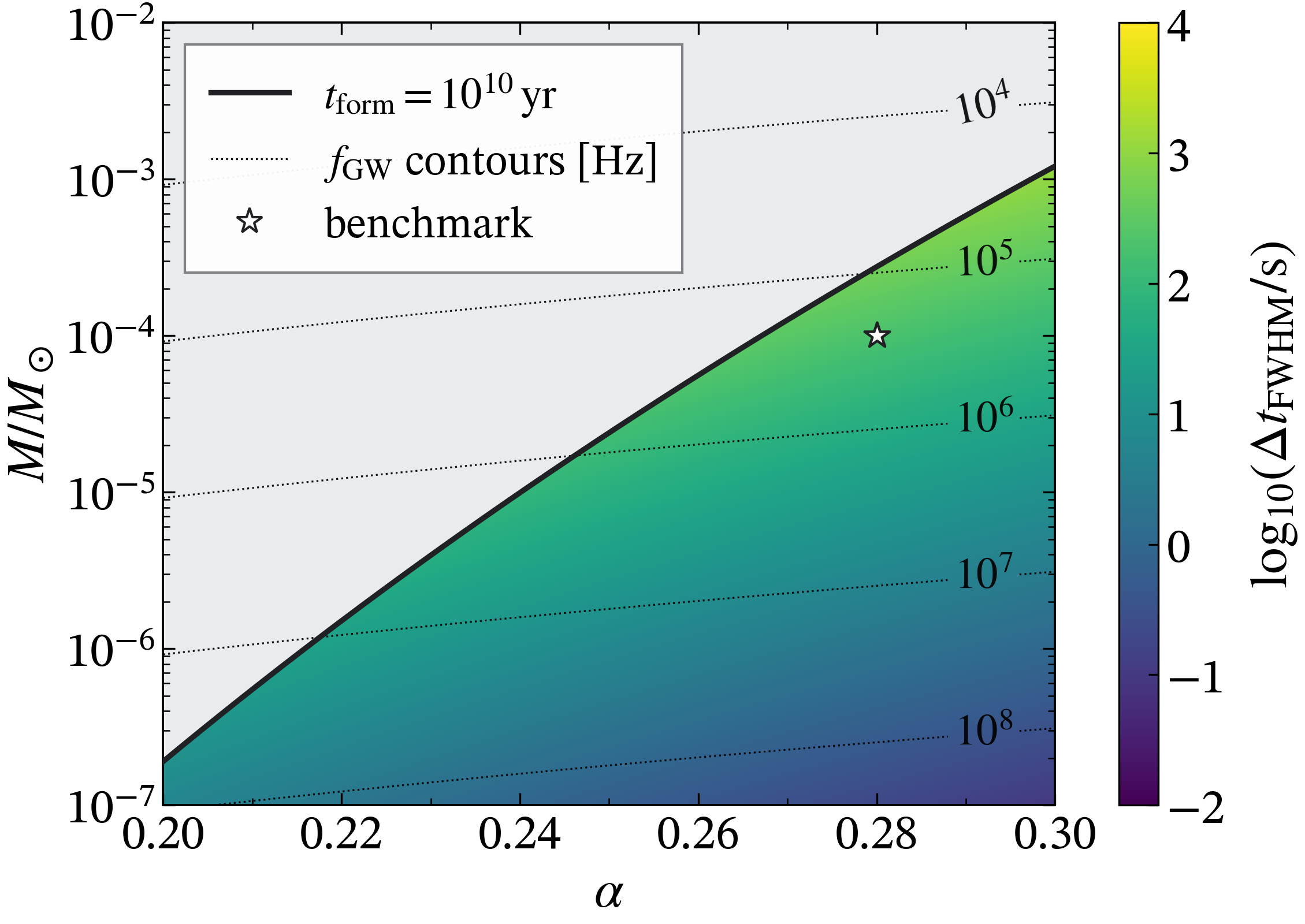}
    \caption{Viable parameter space for the $544\to322$ transition. The coloured wedge satisfies $t_{\rm form}\le10^{10}\,{\rm yr}$, and is shaded by the pulse duration $\log_{10}(\Delta t_{\rm FWHM}/{\rm s})$. The solid curve marks $t_{\rm form}=10^{10}\,{\rm yr}$, the dotted curves are $f_{\rm GW}$ contours, and the star is the benchmark point $(\alpha,M)=(0.28,10^{-4}M_\odot)$.}
    \label{fig:parameter_space}
\end{figure}

The self-stimulated dynamics also determines the outgoing radiation. The macroscopic coherence of the two cloud levels sources a time-dependent mass quadrupole,
\begin{equation}
    M_{ij}(t)=2M_c\,\mathrm{Re}\!\left[\zeta(t)\mathcal Q_{ij}^{21}e^{-i\omega_0t}\right]\,,
    \label{eq:main_quadrupole}
\end{equation}
where $M_c=\sigma M$, $\zeta=c_2^*c_1$, and $\mathcal Q_{ij}^{21}=\langle2|x_ix_j|1\rangle$ follows the convention of Sec.~II. At distance $d$ in the direction $\hat{\mathbf n}$, the far-zone GW is~\cite{Kyriazis:2025fis,Maggiore:2007nxb}
\begin{equation}
    h_{ij}^{\rm TT}(t,d\hat{\mathbf n})
    \simeq\frac{2G}{d}\Lambda_{ij,kl}(\hat{\mathbf n})\ddot M_{kl}(t_{\rm r})\,,
    \label{eq:main_far_zone_projection}
\end{equation}
where $t_{\rm r}=t-d$, $P_{ij}=\delta_{ij}-n_in_j$, and $\Lambda_{ij,kl}=P_{i(k}P_{l)j}-P_{ij}P_{kl}/2$ is the TT projector.

For the two viable transitions in Table~\ref{tab:channel_coefficients}, both with $\Delta m\equiv m_2-m_1=-2$, the emitted GW polarizations take the form
\begin{align}
    h_+(t)&\simeq-h_0\frac{1+\cos^2\iota}{2}
    \operatorname{sech}\!\left[\frac{t_{\rm r}-t_D}{2t_p}\right]\cos(\omega_0t_{\rm r})\,,\nonumber\\
    h_\times(t)&\simeq-h_0\cos\iota\,
    \operatorname{sech}\!\left[\frac{t_{\rm r}-t_D}{2t_p}\right]\sin(\omega_0t_{\rm r})\,,
	\label{eq:main_pulse_waveform}
\end{align}
where $\iota$ is the angle between $\hat{\mathbf n}$ and the spin axis, and
\begin{equation}
    h_0=\frac{2G\sigma M\omega_0^2|Q_{xx}^{21}|}{d}
    \label{eq:main_strain_amplitude}
\end{equation}
is the strain amplitude, with $Q_{xx}^{21}$ the Cartesian $xx$ component of the STF matrix element defined in Sec.~II. The TT projection and polarization formulas are derived in Appendix~\ref{sec:far_zone}. Eq.~\eqref{eq:main_pulse_waveform} is a representative phase-unmodulated result; the additional coherence phase is derived in Appendix~\ref{sec:far_zone}. 

\begin{figure*}[!htbp]
    \centering
    \includegraphics[width=0.8\textwidth]{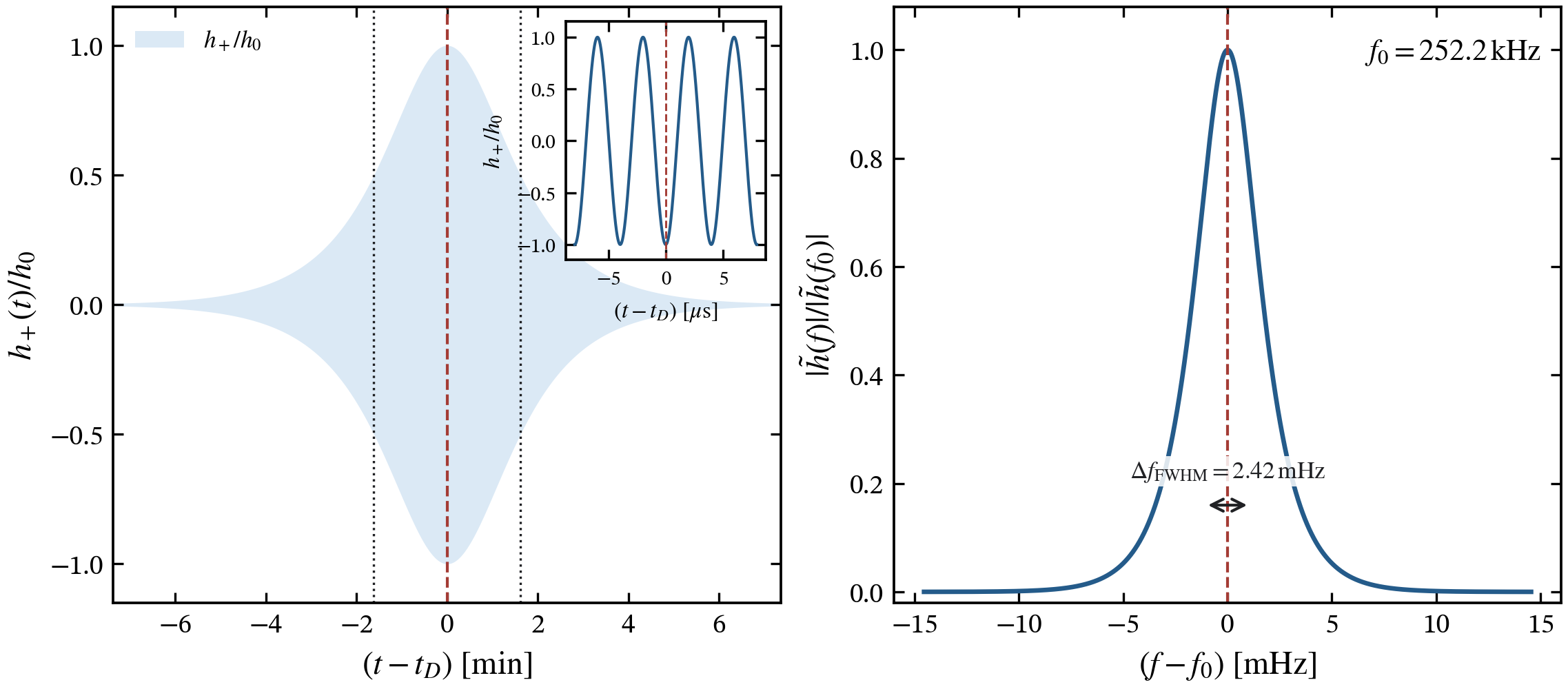}
    \caption{Representative phase-unmodulated waveform and spectrum for the benchmark $544\to322$ signal. Left: the plus-polarization pulse $h_+(t)/h_0$; the $252.2\,{\rm kHz}$ carrier fills the $\operatorname{sech}$ envelope over the $\sim194\,{\rm s}$ duration, with the resolved oscillation around the peak shown in the inset. Right: the analytic lineshape $|\tilde h(f)|/|\tilde h(f_0)|=\operatorname{sech}[2\pi^2t_p(f-f_0)]$, with $f_0$ and the power bandwidth $\Delta f_{\rm FWHM}=2.42\,{\rm mHz}$ marked.}
    \label{fig:waveform}
\end{figure*}

The far-zone strain is proportional to the collective coherence, $h\propto Nc_1c_2^*$, so the same coherence that drives the avalanche also shapes the escaping signal: a quasi-monochromatic pulse with a full width at half-maximum $\Delta t_{\rm FWHM}=4\,\operatorname{arcosh}(2)\,t_p$, and a frequency bandwidth $\Delta f\sim(2\pi t_p)^{-1}$.

The far-zone waveform also provides an independent energy-balance check. Integrating the averaged GW flux over emission directions gives $P_{\rm GW}=N\omega_0\Gamma_{\rm eff}n_1n_2=-\dot E_{\rm cloud}$.  This equality verifies the consistency between the collective transfer rate and the outgoing GW power; a complete population transfer consequently radiates the energy $N\omega_0$. The derivation is given in Appendix~\ref{sec:sm_energy_balance}.

Below, we present the $544\to322$ transition as a representative example. For this channel, the Bohr splitting is $\omega_0=(8/225)\mu\alpha^2$ and the quadrupole matrix element is $Q_{xx}^{21}\simeq24\,r_B^2$. The saturated cloud mass fraction is estimated from sequential superradiant saturation as described in Appendix~\ref{sec:g_cloud_mass}, giving $\sigma\simeq3.33\times10^{-3}$ at $\alpha=0.2$. Eq.~\eqref{eq:main_strain_amplitude} then gives the scaling
\begin{multline}
    h_0\simeq3.9\times10^{-25}
    \left(\frac{\sigma}{3.33\times10^{-3}}\right)
    \left(\frac{\alpha}{0.2}\right)^2\\
    \times\left(\frac{M}{10^{-2}M_\odot}\right)
    \left(\frac{10\,{\rm kpc}}{d}\right)\,.
    \label{eq:main_h0_benchmark}
\end{multline}

We scan the $(M,\alpha)$ parameter space of the $544\to322$ transition following Appendix~\ref{sec:g_cloud_mass}, with the results shown in Fig.~\ref{fig:parameter_space}. The colour bar represents the pulse duration $\log_{10}(\Delta t_{\rm FWHM}/{\rm s})$, with contours of the GW frequency $f_{\rm GW}$ and a benchmark point overlaid. The upper level $\ket{544}$ grows slowly, so its formation within the age of the Universe constrains the viable hosts. We estimate the formation timescale as $t_{\rm form}=100 t_{\rm SR}$, with $t_{\rm SR}=(2\gamma_{544})^{-1}$~\cite{Fernandez:2019qbj,Siemonsen:2022darkphoton}, and require $t_{\rm form}\lesssim10^{10}\,{\rm yr}$, selecting the coloured wedge. Within the scanned range $0.2\le\alpha\le0.3$, the formation-time constraint restricts the host to sub-solar masses, $M\lesssim10^{-3}M_\odot$, so the relevant systems are primordial BHs rather than stellar-origin BHs.

From this region we select the benchmark point $(\alpha,M)=(0.28,\,10^{-4}M_\odot)$ at a distance $d=10\,{\rm kpc}$, with cloud mass fraction $\sigma=6.5\times10^{-3}$ and formation timescale $t_{\rm form}\simeq3.6\times10^9\,{\rm yr}$. The corresponding pulse has transition rate $\Gamma_{\rm eff}=2.7\times10^{-2}\,{\rm s}^{-1}$, duration $\Delta t_{\rm FWHM}=4\operatorname{arcosh}(2)\,t_p\simeq5.27\,t_p\simeq194\,{\rm s}$, frequency $f_0=252.2\,{\rm kHz}$ and strain amplitude $h_0\simeq1.5\times10^{-26}$ from Eq.~\eqref{eq:main_h0_benchmark}. Fig.~\ref{fig:waveform} shows the waveform and its spectrum: rapid carrier oscillations lie within a slowly varying $\operatorname{sech}$ envelope, producing a narrow spectral peak around $f_0$.

We give an illustrative sensitivity estimate using the matched-filter signal-to-noise ratio (SNR), following Appendix~\ref{sec:far_zone}. Approximating the noise spectral density $S_n(f)$ as constant across the narrow signal bandwidth, an observing window centred on the peak gives
\begin{equation}
    {\rm SNR}(T_{\rm obs})=\frac{2h_{\rm eff}}{\sqrt{\Gamma_{\rm eff}S_n(f_0)}}\sqrt{\tanh\left(\frac{T_{\rm obs}}{4t_p}\right)}\,,
    \label{eq:main_pulse_snr}
\end{equation}
where $f_0=\omega_0/(2\pi)$ and $h_{\rm eff}=\mathcal A_{\rm det}h_0$, with $\mathcal A_{\rm det}$ encoding the detector antenna response and source inclination. For the full pulse, the noise amplitude spectral density corresponding to a target SNR $\rho_\star$ is
\begin{equation}
    \sqrt{S_n^{\rm req}(f_0)}=\frac{2\mathcal A_{\rm det}h_0}{\rho_\star\sqrt{\Gamma_{\rm eff}}}\,,
    \label{eq:main_required_asd}
\end{equation}
This estimate provides a sensitivity target for high-frequency GW searches. Assessing observational prospects requires a detector-specific analysis beyond the scope of this work.

\section{{Conclusions}}
We show that superradiant axion clouds can become dynamically unstable under weak resonant external GWs through a cooperative emission process similar to superfluorescence. The incident perturbation seeds a macroscopic two-level coherence in the cloud, activating an interference quadrupole, whose gravitational self-field acts back on the transition. The TT on-shell response supplies the imaginary part of the resonant self-energy that drives population transfer, while the constrained response and TT principal-value contribution are conservative. After the seed has passed, and when lower-level absorption is negligible, this feedback yields logistic population transfer on the timescale $\Gamma_{\rm eff}^{-1}$ and a delayed, quasi-monochromatic GW pulse with a $\operatorname{sech}$ envelope. The outgoing GW power matches the cloud's energy loss, providing an independent consistency check.

Unlike optical superfluorescence, which can exhibit a train of ringing pulses due to spatial propagation and reabsorption~\cite{PhysRevA.14.1169}, the gravitational analogue emits a solitary pulse. Such ringing requires the avalanche timescale to be comparable to the light-crossing time of the medium. For gravitational atoms, this ratio is parametrically small. For a fixed Bohr transition, $\Gamma_{\rm eff} r_B \sim \sigma\alpha^{6} \ll 1$, so the cloud can be treated as responding coherently on the transition timescale, with propagation effects expected to play a limited role in this regime.

Although the self-stimulated transitions require stringent dynamical conditions, favourable channels produce distinctive transient signatures accessible to future high-frequency gravitational-wave detectors. This opens the possibility of probing dark-sector boson clouds through their collective, self-stimulated gravitational emission.
	
	\noindent{\emph{Acknowledgments}}
	We thank Yifan Chen, Yi-kun Li and Cheng-Jun Fang for insightful discussions and comments.
    This work is supported in part by the National Key Research and Development Program of China Grants No. 2020YFC2201501 and No. 2021YFC2203002, in part by the National Natural Science Foundation of China Grants No. 12588101, No. 12235019, No. 12075297 and No. 12147103, in part by the Science Research Grants from the China Manned Space Project with No. CMS-CSST-2021-B01, in part by the Fundamental Research Funds for the Central Universities.

	\newpage
	\begin{titlepage}
		\centering
		{\large\textbf{Supplemental information for ``Self-stimulated avalanche in superradiant axion clouds''} \par}
		\vfill
		$\quad$
	\end{titlepage}    

\appendix
\setcounter{section}{0}

\section{Gravitational atom}
\label{sec:sm_gravitational_atom}

We use natural units, $\hbar=c=1$, and the mostly-plus metric signature.  Consider a real scalar field $\phi$ of mass $\mu$ around a Kerr BH of mass $M$ and angular momentum $J$. In the weak-coupling regime, we define
\begin{equation}
 \phi
 =
 \frac{1}{\sqrt{2\mu}}
 \left(\Psi e^{-i\mu t}+\Psi^*e^{i\mu t}\right)\,.
 \label{eq:sm_nr_decomposition}
\end{equation}
To leading nonrelativistic order, the wave function obeys
\begin{equation}
 i\partial_t\Psi
 =
 \left(-\frac{\nabla^2}{2\mu}-\frac{\alpha}{r}\right)\Psi\,,
 \label{eq:sm_schrodinger}
\end{equation}
where $\alpha\equiv GM\mu\ll1$. The Coulombic equation defines the gravitational Bohr radius
\begin{equation}
 r_B\equiv\frac{1}{\mu\alpha}
 =\frac{GM}{\alpha^2}\,.
 \label{eq:sm_bohr_radius}
\end{equation}

The bound states are labelled by $\lvert n\ell m\rangle$, with
$n\geq1$, $0\leq\ell\leq n-1$, and $-\ell\leq m\leq\ell$. Their
normalised spatial wave functions are
\begin{equation}
 \psi_{n\ell m}(\mathbf x)
 =R_{n\ell}(r)Y_{\ell m}(\theta,\varphi)\,,
 \label{eq:sm_hydrogenic_mode}
\end{equation}
where
\begin{align}
 R_{n\ell}(r)
 ={}&
 \left[
 \left(\frac{2}{nr_B}\right)^3
 \frac{(n-\ell-1)!}{2n(n+\ell)!}
 \right]^{1/2}
 e^{-r/(nr_B)}
 \nonumber\\
 &\times
 \left(\frac{2r}{nr_B}\right)^\ell
 L_{n-\ell-1}^{2\ell+1}
 \left(\frac{2r}{nr_B}\right)\,.
 \label{eq:sm_radial_mode}
\end{align}
Here $Y_{\ell m}$ and $L_p^q$ are the normalised spherical harmonics and generalised Laguerre polynomials. We use
\begin{equation}
 \int d^3\mathbf x\,
 \psi_{n\ell m}^*\psi_{n'\ell'm'}
 =
 \delta_{nn'}\delta_{\ell\ell'}\delta_{mm'}\,.
 \label{eq:sm_mode_normalization}
\end{equation}
The corresponding envelope mode is $\Psi_{n\ell m}=\psi_{n\ell m} e^{-i(\omega_{n\ell m}^{\Psi}-\mu)t}$.
Its characteristic radius and velocity scale as
\begin{equation}
 r\sim n^2r_B\,,
 \qquad
 v\sim\frac{\alpha}{n}\,.
 \label{eq:sm_hydrogenic_scaling}
\end{equation}

Through the order required for Bohr, fine-structure, and hyperfine
transitions, the eigenfrequency is~\cite{Brito:2015oca,Baumann:2019eav}
\begin{equation}
 \omega_{n\ell m}
 =\mu\Bigg[
 1-\frac{\alpha^2}{2n^2}
 -F_{n\ell}\alpha^4
 +h_{n\ell}\tilde a\,m\alpha^5
 +\mathcal O(\alpha^6)
 \Bigg]\,,
 \label{eq:sm_real_spectrum}
\end{equation}
where
\begin{align}
 F_{n\ell}
 &=
 \frac{1}{n^4}
 \left[
 \frac18+\frac{3n-2\ell-1}{\ell+1/2}
 \right]\,,
 \nonumber\\
 h_{n\ell}
 &=
 \frac{16}
 {2n^3\ell(2\ell+1)(2\ell+2)}\,.
 \label{eq:sm_spectrum_coefficients}
\end{align}

The complex mode frequency is separated as
\begin{equation}
 \omega_{n\ell m}^{\Psi}
 =
 \omega_{n\ell m}+i\gamma_{n\ell m}\,.
 \label{eq:sm_complex_mode}
\end{equation}
A positive $\gamma_{n\ell m}$ denotes exponential growth. At leading order
in $\alpha$~\cite{Brito:2015oca,Baumann:2019eav},
\begin{equation}
 \gamma_{n\ell m}
 =
 \frac{2r_+}{GM}
 \left(m\Omega_H-\omega_{n\ell m}\right)
 \alpha^{4\ell+5}
 \mathcal K_{n\ell}\mathcal X_{n\ell m}\,,
 \label{eq:sm_superradiant_rate}
\end{equation}
where
\begin{equation}
 r_+=GM\left(1+\sqrt{1-\tilde a^2}\right)\,,
 \qquad
 \Omega_H=\frac{\tilde a}{2r_+}\,,
 \label{eq:sm_horizon_quantities}
\end{equation}
and
\begin{align}
 \mathcal K_{n\ell}
 ={}&
 \frac{2^{4\ell+1}(n+\ell)!}
 {n^{2\ell+4}(n-\ell-1)!}
 \left[
 \frac{\ell!}{(2\ell)!(2\ell+1)!}
 \right]^2\,,
 \label{eq:sm_K_coefficient}\\
 \mathcal X_{n\ell m}
 ={}&
 \prod_{j=1}^{\ell}
 \left[
 j^2(1-\tilde a^2)
 +4r_+^2
 (m\Omega_H-\omega_{n\ell m})^2
 \right]\,.
 \label{eq:sm_X_coefficient}
\end{align}

\section{Two-level cloud and interaction Hamiltonian}
\label{sec:sm_two_level}

\subsection{Nonrelativistic two-level setup}

The stationary BH field is incorporated into the unperturbed Hamiltonian $H_0$ in Eq.~\eqref{eq:sm_schrodinger}, which determines the bound states. The additional perturbation $h_{\mu\nu}$ denotes only the weak dynamical field sourced by the transition and, when present, by an external seed. For the hydrogenic clouds considered here, we have
\begin{equation}
 v\ll1\,,
 \qquad
 \frac{GM}{n^2r_B}\sim v^2\ll1\,.
 \label{eq:sm_response_hierarchy}
\end{equation}
We therefore use the Minkowski linearized response at the retained order. Background-curvature corrections to this response are relative $\mathcal O(v^2)$; the stationary binding field remains in
$H_0$. Let
\begin{equation}
 H_0\psi_a=\omega_a\psi_a\,,
 \qquad
 \int d^3\mathbf x\,\psi_a^*\psi_b=\delta_{ab}\,,
 \label{eq:sm_unperturbed_modes}
\end{equation}
where $\omega_a=\omega_{n_a\ell_a m_a}-\mu$ is the real binding energy of level $a$. A two-level cloud with occupation number $N$ is described by
\begin{equation}
 \Psi(t,\mathbf x)
 =
 \sqrt N\sum_{a=1,2}
 c_a(t)\psi_a(\mathbf x)
 e^{-i\omega_at}\,,
 \label{eq:sm_two_level_ansatz}
\end{equation}
where
\begin{equation}
 |c_1|^2+|c_2|^2=1\,,
 \qquad
 M_c=N\mu\,.
 \label{eq:sm_cloud_normalization}
\end{equation}
We define the transition frequency $\omega_0$, the occupation fractions $n_a$, and the coherence $\zeta$ by
\begin{equation}
 \omega_0\equiv\omega_1-\omega_2>0\,,
 \quad
 n_a\equiv|c_a|^2\,,
 \quad
 \zeta\equiv c_2^*c_1\,.
 \label{eq:sm_two_level_definitions}
\end{equation}

\subsection{Conserved effective stress tensor}

Substituting the field decomposition~\eqref{eq:sm_nr_decomposition}
into the scalar stress tensor and averaging over the rapid
oscillations removes the terms proportional to
$e^{\pm 2i\mu t}$, leaving
\begin{align}
 T_{\mu\nu}^{\rm cloud}
 ={}&\frac{1}{2\mu}\Big\{
 \partial_\mu\!\left(\Psi^*e^{i\mu t}\right)
 \partial_\nu\!\left(\Psi e^{-i\mu t}\right)
 \nonumber\\
 &+\partial_\nu\!\left(\Psi^*e^{i\mu t}\right)
 \partial_\mu\!\left(\Psi e^{-i\mu t}\right)
 \nonumber\\
 &-\eta_{\mu\nu}\Big[
 \partial^\alpha\!\left(\Psi^*e^{i\mu t}\right)
 \partial_\alpha\!\left(\Psi e^{-i\mu t}\right)+\mu^2\Psi^*\Psi
 \Big]\Big\}\,.
 \label{eq:sm_averaged_cloud_stress}
\end{align}
Substituting the multimode expansion Eq.~\eqref{eq:sm_two_level_ansatz} into the averaged scalar stress tensor gives the cloud contribution
\begin{equation}
 T_{\mu\nu}^{\rm cloud}
 \simeq
 M_c\sum_{a,b}
 c_a^*c_b\,
 \tau_{\mu\nu}^{ab,{\rm cloud}}(\mathbf x)e^{i(\omega_a-\omega_b)t}\,,
 \label{eq:sm_cloud_stress}
\end{equation}
where
\begin{align}
\tau_{00}^{ab,\mathrm{cloud}}
={}&
\frac{1}{2\mu^2}\Big\{
\left[(\mu+\omega_a)(\mu+\omega_b)+\mu^2\right]
\nonumber\\
&\qquad\times\psi_a^*\psi_b
+\partial_k\psi_a^*\partial_k\psi_b
\Big\}\,,\label{eq:sm_cloud_density_coefficient}\\
\tau_{0i}^{ab,\mathrm{cloud}}
={}&
\frac{i}{2\mu^2}
\Big[
(\mu+\omega_a)\psi_a^*\partial_i\psi_b
\nonumber\\
&\qquad-(\mu+\omega_b)\left(\partial_i\psi_a^*\right)\psi_b
\Big]\,,\label{eq:sm_cloud_momentum_coefficient}\\
\tau_{ij}^{ab,\mathrm{cloud}}
={}&
\frac{1}{2\mu^2}
\Big[
\partial_i\psi_a^*\partial_j\psi_b
+\partial_j\psi_a^*\partial_i\psi_b
-\delta_{ij}\partial_k\psi_a^*\partial_k\psi_b\nonumber\\
&\hspace{4.7em}
+\delta_{ij}
\left[(\mu+\omega_a)(\mu+\omega_b)-\mu^2\right]
\nonumber\\
&\hspace{4.7em}\times
\psi_a^*\psi_b
\Big]\,.
\label{eq:sm_cloud_spatial_stress}
\end{align}
Here we neglect the time derivatives of the slowly varying envelopes $c_a(t)$; the factors $\mu+\omega_a$ arise from differentiating the full positive-frequency field $\Psi e^{-i\mu t}$.

The cloud exchanges momentum with the stationary binding field, so $T_{\mu\nu}^{\rm cloud}$ alone is not a conserved flat-background source. We include the cloud-dependent binding contribution and define
\begin{equation}
 \tau_{\mu\nu}^{ab}
 \equiv
 \tau_{\mu\nu}^{ab,{\rm cloud}}
 +\tau_{\mu\nu}^{ab,{\rm bind}}\,.
 \label{eq:sm_complete_transition_tensor}
\end{equation}
At the order retained, the binding response is linear in the cloud's quadratic source and therefore carries the same mode bilinears $c_a^*c_b$. Since the binding background is stationary, each contribution retains its difference frequency $\omega_a-\omega_b$, giving the complete effective source
\begin{equation}
 T_{\mu\nu}(t,\mathbf x)
 \simeq M_c\sum_{a,b}
 c_a^*c_b\,
 \tau_{\mu\nu}^{ab}(\mathbf x)e^{i(\omega_a-\omega_b)t}\,.
 \label{eq:sm_complete_stress}
\end{equation}
At this order, the complete effective source satisfies energy--momentum conservation, $\partial^\mu T_{\mu\nu}=0$. In particular, the resonant two-level sectors obey
\begin{equation}
 \partial^\mu
 \left(\tau_{\mu\nu}^{21}e^{-i\omega_0t}\right)=0\,,
 \quad
 \partial^\mu
 \left(\tau_{\mu\nu}^{12}e^{i\omega_0t}\right)
 =0\,,
 \label{eq:sm_transition_conservation}
\end{equation}
and
\begin{equation}
 \tau_{\mu\nu}^{ba}
 =\left(\tau_{\mu\nu}^{ab}\right)^*\,.
 \label{eq:sm_transition_hermiticity}
\end{equation}
Expanding the sum in Eq.~\eqref{eq:sm_complete_stress} for $a,b=1,2$ gives
\begin{align}
 T_{\mu\nu}\simeq M_c\big[
 &n_1\tau_{\mu\nu}^{11}
 +n_2\tau_{\mu\nu}^{22}
 \nonumber\\
 &+\zeta\tau_{\mu\nu}^{21}e^{-i\omega_0t}
 +\zeta^*\tau_{\mu\nu}^{12}e^{i\omega_0t}
 \big]\,.
 \label{eq:sm_two_level_stress}
\end{align}

\subsection{Interaction Hamiltonian and tidal limit}

To first order in the dynamical metric perturbation, the matter action is $S_{\rm int}^{(1)}=\frac12\int dt\,d^3\mathbf x\,h_{\mu\nu}T^{\mu\nu}$ and the interaction Hamiltonian is
\begin{equation}
 H_{\rm int}^{(1)}
 =
 -\frac12\int d^3\mathbf x\,
 h^{\mu\nu}T_{\mu\nu}\,.
 \label{eq:sm_linear_interaction}
\end{equation}
Substitution of Eq.~\eqref{eq:sm_complete_stress} gives
\begin{equation}
 H_{\rm int}^{(1)}
 =
 N\sum_{a,b}
 c_a^*c_b\,e^{i(\omega_a-\omega_b)t}
 V_{ab}(t)\,,
 \label{eq:sm_projected_hamiltonian}
\end{equation}
with the complete transition vertex
\begin{equation}
 V_{ab}(t)
 =
 -\frac{\mu}{2}
 \int d^3\mathbf x\,
 \tau_{\mu\nu}^{ab}(\mathbf x)
 h^{\mu\nu}(t,\mathbf x)\,.
 \label{eq:sm_complete_vertex}
\end{equation}
The resulting two-level equations are
\begin{align}
 i\dot c_1
 &=V_{11}c_1+V_{12}e^{i\omega_0t}c_2\,,
 \nonumber\\
 i\dot c_2
 &=V_{22}c_2+V_{21}e^{-i\omega_0t}c_1\,.
 \label{eq:sm_two_level_equations}
\end{align}
For a real perturbation,
$V_{ba}=V_{ab}^*$ follows from
Eq.~\eqref{eq:sm_transition_hermiticity}.

In comoving Fermi coordinates $(t,\mathbf{x})$, whose origin is located at the centre-of-mass of the unperturbed BH-cloud, the leading spatial expansion is
\begin{equation}
 h_{00}(t,\mathbf x)
 =
 -R_{0i0j}(t,\mathbf0)x_ix_j+\mathcal O(x^3)\,,
 \label{eq:sm_fermi_metric}
\end{equation}
where the linearized curvature is
\begin{equation}
 R_{0i0j}
 =\frac12\big(
 \partial_i\dot h_{0j}
 +\partial_j\dot h_{0i}
 -\partial_i\partial_jh_{00}
 -\ddot h_{ij}\big)\,.
 \label{eq:sm_linearized_riemann}
\end{equation}
Following the nonrelativistic reduction used in Sec.~II, the metric perturbation produces the leading tidal potential
\begin{equation}
 V_{\rm tidal}(t,\mathbf x)
 =-\frac{\mu}{2}h^{00}(t,\mathbf x)
 \simeq\frac{\mu}{2}E_{ij}(t)x_ix_j\,,
 \label{eq:sm_tidal_potential}
\end{equation}
where $E_{ij}(t)\equiv R_{0i0j}(t,\mathbf0)$. Projection onto the bound states gives
\begin{equation}
 V_{ab}
 \simeq
 \frac{\mu}{2}E_{ij}(t)\mathcal Q_{ij}^{ab}\,,
 \label{eq:sm_tidal_vertex}
\end{equation}
where
\begin{equation}
 E_{ij}(t)\equiv R_{0i0j}(t,\mathbf0)\,,
 \qquad
 \mathcal Q_{ij}^{ab}
 \equiv
 \int d^3\mathbf x\,
 \psi_a^*x_ix_j\psi_b\,.
\label{eq:sm_tidal_definitions}
\end{equation}

To check consistency with the complete vertex Eq.~\eqref{eq:sm_complete_vertex}, define the source
quadrupole per unit cloud mass by
\begin{equation}
 \mathcal I_{ij}^{ab}
 \equiv\int d^3\mathbf x\,x_ix_j\tau_{00}^{ab}(\mathbf x)\,.
 \label{eq:sm_complete_source_quadrupole}
\end{equation}
Its cloud contribution is
$\mathcal I_{ij}^{ab,\mathrm{cloud}}
=\mathcal Q_{ij}^{ab}+\delta\mathcal I_{ij}^{ab,\mathrm{cloud}}$.
Eq.~\eqref{eq:sm_cloud_density_coefficient} gives explicitly
\begin{align}
 \delta\mathcal I_{ij}^{ab,\mathrm{cloud}}
 ={}&\frac{(\mu+\omega_a)(\mu+\omega_b)-\mu^2}{2\mu^2}
 \mathcal Q_{ij}^{ab}\nonumber\\
 &+\frac{1}{2\mu^2}\int d^3\mathbf x\,
 x_ix_j\partial_k\psi_a^*\partial_k\psi_b\,.
 \label{eq:sm_cloud_quadrupole_correction}
\end{align}
Let $L$ and $v$ denote the characteristic size and orbital velocity of the participating bound states. Their binding energies are $\omega_a=\mathcal O(\mu v^2)$ and their characteristic momenta are $\mathcal O(\mu v)$. Thus $[(\mu+\omega_a)(\mu+\omega_b)-\mu^2]/(2\mu^2)=\mathcal O(v^2)$. Normalisation gives $\mathcal Q_{ij}^{ab}=\mathcal O(L^2)$ and $\int d^3\mathbf x\,x_ix_j\partial_k\psi_a^*\partial_k\psi_b=\mathcal O(\mu^2v^2L^2)$, so both terms in Eq.~\eqref{eq:sm_cloud_quadrupole_correction} are $\mathcal O(v^2L^2)$.

In the weak-field source-multipole expansion, the cloud-dependent binding and gravitational field energies enter at order $M_cv^2$, relative to the leading rest-mass contribution $M_c$. Their quadrupole correction at the source scale $L$ therefore has the power counting
\begin{equation}
 \mathcal I_{ij}^{ab,\mathrm{bind}}
 \equiv\int d^3\mathbf x\,x_ix_j\tau_{00}^{ab,\mathrm{bind}}
 =\mathcal O(v^2L^2)\,.
 \label{eq:sm_binding_quadrupole_correction}
\end{equation}
Combining the two contributions,
\begin{align}
 \mathcal I_{ij}^{ab}
 &=\mathcal Q_{ij}^{ab}+\delta\mathcal I_{ij}^{ab}\,,\nonumber\\
 \delta\mathcal I_{ij}^{ab}
 &\equiv\delta\mathcal I_{ij}^{ab,\mathrm{cloud}}
 +\mathcal I_{ij}^{ab,\mathrm{bind}}
 =\mathcal O(v^2L^2)\,.
 \label{eq:sm_source_multipole_matching}
\end{align}
For an unsuppressed quadrupole, this is a relative
$\mathcal O(v^2)=\mathcal O(\alpha^2/n^2)$ correction, using Eq.~\eqref{eq:sm_hydrogenic_scaling}.
The leading electric-quadrupole part of the complete vertex is thus
\begin{equation}
 V_{ab}
 \simeq\frac{\mu}{2}E_{ij}\mathcal I_{ij}^{ab}
 \simeq\frac{\mu}{2}E_{ij}\mathcal Q_{ij}^{ab}\,,
 \label{eq:sm_tidal_consistency}
\end{equation}
in agreement with the direct tidal reduction above.
We retain the complete vertex \eqref{eq:sm_complete_vertex} for the
constraint response and the principal-value TT response.  In the
self-energy calculation, the quadrupole reduction is applied only
to the radiative TT shell.

\section{SVT response of the transition field}
\label{sec:sm_svt_response}

\subsection{Metric decomposition and gauge-invariant variables}

The spatial SVT decomposition of the metric perturbation is
\begin{align}
 h_{00}&=2A\,,
 \nonumber\\
 h_{0i}&=\beta_i^{\rm T}+\partial_i\gamma\,,
 \nonumber\\
 h_{ij}
 &=h_{ij}^{\rm TT}
 +2\partial_{(i}F_{j)}^{\rm T}
 \nonumber\\
 &\quad+
 \left(\partial_i\partial_j
 -\frac13\delta_{ij}\nabla^2\right)B
 +\frac13\delta_{ij}\mathcal H\,.
 \label{eq:sm_metric_svt}
\end{align}
Here $A$, $\gamma$, $B$, and $\mathcal H$ are scalars, $\beta_i^{\rm T}$ and $F_i^{\rm T}$ are transverse vectors, and $h_{ij}^{\rm TT}$ is transverse and traceless:
\begin{equation}
 \partial_i\beta_i^{\rm T}
 =
 \partial_iF_i^{\rm T}
 =
 \partial_i h_{ij}^{\rm TT}
 =
 h_{ii}^{\rm TT}=0\,.
 \label{eq:sm_svt_conditions}
\end{equation}
The gauge-invariant combinations are
\begin{align}
 \Phi
 &\equiv
 \dot\gamma-\frac12h_{00}-\frac12\ddot B\,,
 \nonumber\\
 \Theta
 &\equiv
 \frac13(\mathcal H-\nabla^2B)\,,
 \nonumber\\
 \Xi_i
 &\equiv
 \beta_i^{\rm T}-\dot F_i^{\rm T}\,.
 \label{eq:sm_gauge_invariants}
\end{align}
Direct substitution into Eq.~\eqref{eq:sm_linearized_riemann} gives
\begin{equation}
 R_{0i0j}
 =-\frac12\ddot h_{ij}^{\rm TT}+\partial_i\partial_j\Phi-\frac12\delta_{ij}\ddot\Theta+\partial_{(i}\dot\Xi_{j)}\,.
 \label{eq:sm_tidal_svt}
\end{equation}

\subsection{Separated field equations}

At the order specified by Eq.~\eqref{eq:sm_response_hierarchy}, the dynamical perturbation obeys
\begin{equation}
 G_{\mu\nu}^{(1)}[h]=8\pi G T_{\mu\nu}
 \label{eq:sm_linearized_einstein}
\end{equation}
on the Minkowski background. Define the source projections
\begin{equation}
 T_{00}\equiv\rho\,,
 \qquad
 T_{0i}\equiv\partial_i u+u_i^{\rm T}\,,
 \qquad
 3p\equiv T_{ii}\,,
 \label{eq:sm_source_svt}
\end{equation}
with $\partial_i u_i^{\rm T}=0$, and
$\Pi_{ij}^{\rm TT}\equiv(T_{ij})^{\rm TT}$.
The required projections of the linearized Einstein tensor are
\begin{align}
 G_{00}^{(1)}
 &=-\nabla^2\Theta\,,
 \nonumber\\
 G_{0i}^{(1)}
 &=-\frac12\nabla^2\Xi_i-\partial_i\dot\Theta\,,
 \nonumber\\
 G_{kk}^{(1)}
 &=\nabla^2(2\Phi+\Theta)-3\ddot\Theta\,,
 \nonumber\\
 (G_{ij}^{(1)})^{\rm TT}
 &=-\frac12\Box h_{ij}^{\rm TT}\,,
 \label{eq:sm_einstein_svt}
\end{align}
where $\Box=-\partial_t^2+\nabla^2$.  Their scalar, vector, and tensor
parts yield
\begin{align}
 \nabla^2\Theta
 &=-8\pi G\rho\,,
 \nonumber\\
 \nabla^2\Phi
 &=4\pi G(\rho+3p-3\dot u)\,,
 \nonumber\\
 \nabla^2\Xi_i
 &=-16\pi G u_i^{\rm T}\,,
 \nonumber\\
 \Box h_{ij}^{\rm TT}
 &=-16\pi G\Pi_{ij}^{\rm TT}\,.
 \label{eq:sm_separated_field_equations}
\end{align}
The first three equations in
Eq.~\eqref{eq:sm_separated_field_equations} are Poisson constraints;
only the TT equation is hyperbolic.

\subsection{Fourier response to the resonant transition}

We use
\begin{equation}
 X(t,\mathbf x)
 =\int\frac{d\omega}{2\pi}\frac{d^3\mathbf k}{(2\pi)^3}\widetilde X(\omega,\mathbf k)e^{-i\omega t+i\mathbf k\cdot\mathbf x}\,.
 \label{eq:sm_fourier_convention}
\end{equation}
Here $\omega$ and $\mathbf k$ are the Fourier frequency and spatial wave vector, and a tilde denotes a Fourier coefficient. A real field satisfies $\widetilde X(-\omega,-\mathbf k)=\widetilde X(\omega,\mathbf k)^*$.
Let
\begin{equation}
 q\equiv|\mathbf k|\,,
 \qquad
 \hat k_i\equiv k_i/q\,,
 \qquad
 P_{ij}\equiv\delta_{ij}-\hat k_i\hat k_j\,.
 \label{eq:sm_transverse_projector}
\end{equation}
For $q\neq0$, the TT projector is
\begin{equation}
 \Lambda_{ij,kl}
 \equiv
 P_{i(k}P_{l)j}-\frac12P_{ij}P_{kl}\,.
 \label{eq:sm_tt_projector}
\end{equation}

We now Fourier transform the effective two-level stress tensor in
Eq.~\eqref{eq:sm_two_level_stress}. In evaluating its response at time
$t_\star$, we hold the slowly varying populations and coherence fixed,
with $\zeta_\star=\zeta(t_\star)$. Define the spatial transition
coefficients by
\begin{equation}
 \widetilde\tau_{\mu\nu}^{ab}(\mathbf k)
 \equiv
 \int d^3\mathbf x\,
 e^{-i\mathbf k\cdot\mathbf x}
 \tau_{\mu\nu}^{ab}(\mathbf x)\,.
 \label{eq:sm_transition_fourier}
\end{equation}
The diagonal terms have zero frequency in this approximation. The
resonant components are
\begin{align}
 \widetilde T_{\mu\nu}^{(+)}
 &=
 2\pi M_c\zeta_\star
 \delta(\omega-\omega_0)
 \widetilde\tau_{\mu\nu}^{21}(\mathbf k)\,,
 \nonumber\\
 \widetilde T_{\mu\nu}^{(-)}
 &=
 2\pi M_c\zeta_\star^*
 \delta(\omega+\omega_0)
 \widetilde\tau_{\mu\nu}^{12}(\mathbf k)\,.
 \label{eq:sm_resonant_fourier_source}
\end{align}
Here $(+)$ and $(-)$ denote support at
$\omega=\pm\omega_0$, respectively, and $\delta$ is the Dirac delta
distribution.  Moreover,
\begin{equation}
 \widetilde\tau_{\mu\nu}^{12}(\mathbf k)
 =
 \left[
 \widetilde\tau_{\mu\nu}^{21}(-\mathbf k)
 \right]^*\,.
 \label{eq:sm_fourier_hermiticity}
\end{equation}

To apply the separated field equations~\eqref{eq:sm_separated_field_equations}
to the positive-frequency transition source, define
$\mathcal A_+(\omega)=2\pi M_c\zeta_\star\delta(\omega-\omega_0)$.
Both the source projections and their sourced gauge-invariant fields
carry this common distribution. We denote the remaining coefficients
by the superscript $21$:
\begin{align*}
 \widetilde X^{(+)}(\omega,\mathbf k)
 &=\mathcal A_+(\omega)\widetilde X^{21}(\mathbf k)\,,\\
 X&\in\{\rho,u,u_i^{\rm T},p,\Pi_{ij}^{\rm TT},\\[-0.3em]
 &\hspace{2.5em}\Phi,\Theta,\Xi_i,h_{ij}^{\rm TT}\}\,.
\end{align*}
Comparing with the source projections defined in
Eq.~\eqref{eq:sm_source_svt} gives
\begin{align}
 \widetilde\rho^{21}
 &\equiv\widetilde\tau_{00}^{21}\,,
 \nonumber\\
 \widetilde u^{21}
 &\equiv-\frac{i}{q^2}
 k_l\widetilde\tau_{0l}^{21}\,,
 \nonumber\\
 \widetilde u_i^{{\rm T},21}
 &\equiv P_{il}\widetilde\tau_{0l}^{21}\,,
 \nonumber\\
 3\widetilde p^{21}
 &\equiv\widetilde\tau_{ii}^{21}\,,
 \nonumber\\
 \widetilde\Pi_{ij}^{{\rm TT},21}
 &\equiv
 \Lambda_{ij,kl}\widetilde\tau_{kl}^{21}\,.
 \label{eq:sm_transition_projections}
\end{align}
Stress-tensor conservation gives
\begin{equation}
 q^2\widetilde u^{21}
 =i\omega_0\widetilde\rho^{21}\,,
 \qquad
 k_i\widetilde\tau_{ij}^{21}
 =-\omega_0\widetilde\tau_{0j}^{21}\,.
 \label{eq:sm_fourier_conservation}
\end{equation}
Using $\partial_t\to-i\omega$ and $\nabla^2\to-q^2$ in Eq.~\eqref{eq:sm_separated_field_equations}, and imposing the retarded prescription $+i0\equiv i\epsilon$ with $\epsilon\to0^+$ on the TT sector, gives
\begin{align}
 \widetilde\Theta^{21}
 &=
 \frac{8\pi G}{q^2}\widetilde\rho^{21}\,,
 \nonumber\\
 \widetilde\Phi^{21}
 &=
 -\frac{4\pi G}{q^2}
 \left(
 \widetilde\rho^{21}
 +3\widetilde p^{21}
 +3i\omega_0\widetilde u^{21}
 \right)\,,
 \nonumber\\
 \widetilde\Xi_i^{21}
 &=
 \frac{16\pi G}{q^2}
 \widetilde u_i^{{\rm T},21}\,,
 \label{eq:sm_constraint_response}
\end{align}
and
\begin{equation}
 \widetilde h_{ij}^{{\rm TT},21}
 =
 \frac{16\pi G}
 {q^2-(\omega_0+i0)^2}
 \Lambda_{ij,kl}(\hat{\mathbf k})
 \widetilde\tau_{kl}^{21}(\mathbf k)\,.
 \label{eq:sm_tt_response}
\end{equation}
The first three responses contain only inverse spatial Laplacians, whereas the TT response contains the propagating pole.

To reconstruct the metric entering the complete interaction vertex~\eqref{eq:sm_complete_vertex}, we choose $B=\gamma=0$ and $F_i^{\rm T}=0$. In this gauge,
\begin{equation}
 h_{00}=-2\Phi\,,
 \qquad
 h_{0i}=\Xi_i\,,
 \qquad
 h_{ij}=\delta_{ij}\Theta+h_{ij}^{\rm TT}\,.
 \label{eq:sm_convenient_gauge}
\end{equation}

The positive-frequency Fourier components are therefore
\begin{align}
 \widetilde h_{00}^{(+)}(\omega,\mathbf k)
 &=-2\mathcal A_+(\omega)\widetilde\Phi^{21}(\mathbf k)\,,
 \nonumber\\
 \widetilde h_{0i}^{(+)}(\omega,\mathbf k)
 &=\mathcal A_+(\omega)\widetilde\Xi_i^{21}(\mathbf k)\,,
 \nonumber\\
 \widetilde h_{ij}^{(+)}(\omega,\mathbf k)
 &=\mathcal A_+(\omega)\Big[
 \delta_{ij}\widetilde\Theta^{21}(\mathbf k)
 +\widetilde h_{ij}^{{\rm TT},21}(\mathbf k)
 \Big]\,.
 \label{eq:sm_fourier_metric_reconstruction}
\end{align}
After the inverse frequency transform and restoring
$\zeta_\star\to\zeta(t)$, the positive-frequency self-field is
\begin{align}
 h_{00}^{(+)}
 ={}&
 -2M_c\zeta(t)e^{-i\omega_0t}
 \int\frac{d^3\mathbf k}{(2\pi)^3}
 \widetilde\Phi^{21}e^{i\mathbf k\cdot\mathbf x}\,,
 \nonumber\\
 h_{0i}^{(+)}
 ={}&
 M_c\zeta(t)e^{-i\omega_0t}
 \int\frac{d^3\mathbf k}{(2\pi)^3}
 \widetilde\Xi_i^{21}e^{i\mathbf k\cdot\mathbf x}\,,
 \nonumber\\
 h_{ij}^{(+)}
 ={}&
 M_c\zeta(t)e^{-i\omega_0t}
 \int\frac{d^3\mathbf k}{(2\pi)^3}
 \nonumber\\[-0.3em]
 &\times
 \left(
 \delta_{ij}\widetilde\Theta^{21}
 +\widetilde h_{ij}^{{\rm TT},21}
 \right)e^{i\mathbf k\cdot\mathbf x}\,.
 \label{eq:sm_positive_frequency_metric}
\end{align}
The negative-frequency field is its complex conjugate.  The inverse Laplacians are defined with decaying boundary conditions at spatial infinity. The diagonal $\omega=0$ response is absorbed into the conservative level energies.

\section{Self-energy and dissipative dynamics}
\label{sec:sm_self_energy}

The positive-frequency self-field in Eq.~\eqref{eq:sm_positive_frequency_metric} enters the complete vertex as
\begin{equation}
 V_{12}^{(+),{\rm self}}(t)
 =-\frac{\mu}{2}
 \int d^3\mathbf x\,
 \tau_{\mu\nu}^{12}(\mathbf x)
 h^{(+)\,\mu\nu}(t,\mathbf x)\,.
 \label{eq:sm_self_energy_vertex}
\end{equation}
Using the spatial Fourier convention,
\begin{equation}
 \int d^3\mathbf x\,
 \tau_{\mu\nu}^{12}(\mathbf x)e^{i\mathbf k\cdot\mathbf x}
 =\widetilde\tau_{\mu\nu}^{12}(-\mathbf k)\,.
 \label{eq:sm_vertex_fourier_integral}
\end{equation}
Substituting the inverse-transform fields from
Eq.~\eqref{eq:sm_positive_frequency_metric} and interchanging the
$\mathbf x$ and $\mathbf k$ integrals therefore gives
\begin{align}
 V_{12}^{(+),{\rm self}}(t)
 ={}&-\frac{\mu M_c}{2}\zeta(t)e^{-i\omega_0t}
 \int\frac{d^3\mathbf k}{(2\pi)^3}
 \nonumber\\
 &\times\Big[
 -2\widetilde\rho^{12}(-\mathbf k)
 \widetilde\Phi^{21}(\mathbf k)
 \nonumber\\
 &\qquad-2\widetilde\tau_{0i}^{12}(-\mathbf k)
 \widetilde\Xi_i^{21}(\mathbf k)
 \nonumber\\
 &\qquad+3\widetilde p^{12}(-\mathbf k)
 \widetilde\Theta^{21}(\mathbf k)
 \nonumber\\
 &\qquad+\widetilde\tau_{ij}^{12}(-\mathbf k)
 \widetilde h_{ij}^{{\rm TT},21}(\mathbf k)
 \Big]\,.
 \label{eq:sm_self_field_vertex_fourier}
\end{align}
Factoring out the common envelope and carrier, we define the resonant self-energy by
\begin{equation}
 V_{12}^{(+),{\rm self}}(t)
 \equiv
 \frac12\Sigma_{12}
 \zeta(t)e^{-i\omega_0t}\,,
 \label{eq:sm_self_energy_definition}
\end{equation}
and decompose it as
\begin{equation}
 \Sigma_{12}
 =
 \Sigma_{{\rm C},12}
 +\Sigma_{{\rm TT},12}\,.
 \label{eq:sm_self_energy_decomposition}
\end{equation}

\subsection{Conservative constraint sector}

The first three terms of Eq.~\eqref{eq:sm_self_field_vertex_fourier} give
\begin{align}
 \Sigma_{{\rm C},12}
 =-\mu M_c\int\frac{d^3\mathbf k}{(2\pi)^3}
 \big[
 &-2\widetilde\rho^{12}(-\mathbf k)
 \widetilde\Phi^{21}(\mathbf k)
 \nonumber\\
 &-2\widetilde\tau_{0i}^{12}(-\mathbf k)
 \widetilde\Xi_i^{21}(\mathbf k)
 \nonumber\\
 &+3\widetilde p^{12}(-\mathbf k)
 \widetilde\Theta^{21}(\mathbf k)
 \big]\,.
 \label{eq:sm_constraint_self_energy_raw}
\end{align}
Only the transverse part of $\widetilde\tau_{0i}^{12}$ contributes to the second line. Substituting the constraint responses~\eqref{eq:sm_constraint_response} and using the conjugation and conservation relations~\eqref{eq:sm_fourier_hermiticity} and~\eqref{eq:sm_fourier_conservation}, this becomes
\begin{align}
 \Sigma_{{\rm C},12}
 ={}&
 -8\pi G\mu M_c
 \int\frac{d^3\mathbf k}{(2\pi)^3q^2}
 \Bigg[
 \left(1-\frac{3\omega_0^2}{q^2}\right)
 |\widetilde\rho^{21}|^2
 \nonumber\\
 &+3\left(
 \widetilde\rho^{21*}\widetilde p^{21}
 +\widetilde p^{21*}\widetilde\rho^{21}
 \right)-4\widetilde u_i^{{\rm T},21*}
 \widetilde u_i^{{\rm T},21}
 \Bigg]\,.
 \label{eq:sm_constraint_self_energy}
\end{align}
The integrand in Eq.~\eqref{eq:sm_constraint_self_energy} is real and
contains no radiative prescription. Therefore
\begin{equation}
 \operatorname{Im}\Sigma_{{\rm C},12}=0\,.
 \label{eq:sm_constraint_no_imaginary}
\end{equation}
The constraint sector contributes only a conservative level shift.

\subsection{TT on-shell contribution}

Substitution of Eq.~\eqref{eq:sm_tt_response} into
the last term of Eq.~\eqref{eq:sm_self_field_vertex_fourier} yields
\begin{align}
 \Sigma_{{\rm TT},12}
 ={}&
 -16\pi G\mu M_c
 \int\frac{d^3\mathbf k}{(2\pi)^3}
 \frac{\widetilde\tau_{ij}^{12}(-\mathbf k)}
 {q^2-(\omega_0+i0)^2}
 \nonumber\\
 &\times
 \Lambda_{ij,kl}(\hat{\mathbf k})
 \widetilde\tau_{kl}^{21}(\mathbf k)\,.
 \label{eq:sm_tt_self_energy}
\end{align}
The numerator is a Hermitian quadratic form.  For $\omega_0>0$,
\begin{equation}
 \frac{1}{q^2-(\omega_0+i0)^2}
 =
 \operatorname{PV}\frac{1}{q^2-\omega_0^2}
 +i\pi\delta(q^2-\omega_0^2)\,,
 \label{eq:sm_retarded_decomposition}
\end{equation}
where $\operatorname{PV}$ denotes the Cauchy principal value.  The
principal-value part is conservative, while the imaginary part is
supported on the radiative shell:
\begin{align}
 \operatorname{Im}\Sigma_{{\rm TT},12}
 ={}&
 -16\pi^2G\mu M_c
 \int\frac{d^3\mathbf k}{(2\pi)^3}
 \delta(q^2-\omega_0^2)
 \nonumber\\
 &\times
 \widetilde\tau_{ij}^{12}(-\mathbf k)
 \Lambda_{ij,kl}
 \widetilde\tau_{kl}^{21}(\mathbf k)\,.
 \label{eq:sm_tt_cut}
\end{align}

Conservation of the $21$ transition implies
\begin{equation}
 \partial_k\tau_{k0}^{21}
 =-i\omega_0\tau_{00}^{21}\,,
 \qquad
 \partial_k\tau_{kj}^{21}
 =-i\omega_0\tau_{0j}^{21}\,,
 \label{eq:sm_position_conservation}
\end{equation}
and hence
$\partial_k\partial_l\tau_{kl}^{21}
=-\omega_0^2\tau_{00}^{21}$.
Using the complete source quadrupole defined in
Eq.~\eqref{eq:sm_complete_source_quadrupole}, two integrations by
parts give the exact zero-spatial-momentum
identity
\begin{equation}
 \widetilde\tau_{ij}^{21}(\mathbf k=\mathbf0)
 =
 -\frac{\omega_0^2}{2}\mathcal I_{ij}^{21}\,.
 \label{eq:sm_zero_momentum_identity}
\end{equation}

The source-multipole matching established in
Eq.~\eqref{eq:sm_source_multipole_matching} gives
\begin{equation}
 \mathcal I_{ij}^{21}
 =
 \mathcal Q_{ij}^{21}
 +\delta\mathcal I_{ij}^{21}\,,
 \qquad
 \delta\mathcal I_{ij}^{21}
 =\mathcal O(v^2L^2)\,,
 \label{eq:sm_quadrupole_matching}
\end{equation}
with $L$ and $v$ defined in the derivation of that matching relation. The spatial form factor has the expansion
\begin{align}
 \widetilde\tau_{ij}^{21}(\mathbf k)
 ={}&
 \widetilde\tau_{ij}^{21}(\mathbf0)
 \nonumber\\
 &-ik_m\int d^3\mathbf x\,
 x_m\tau_{ij}^{21}(\mathbf x)+\cdots\,.
 \label{eq:sm_form_factor}
\end{align}
On the radiative shell, $\mathbf k=\omega_0\hat{\mathbf k}$.
For $\omega_0L\ll1$, Eqs.~\eqref{eq:sm_zero_momentum_identity}--%
\eqref{eq:sm_form_factor} yield
\begin{equation}
 \widetilde\tau_{ij}^{21}
 (\omega_0\hat{\mathbf k})
 \simeq
 -\frac{\omega_0^2}{2}\mathcal Q_{ij}^{21}\,.
 \label{eq:sm_on_shell_matching}
\end{equation}
The omitted terms are relative $\mathcal O(v^2)$ and
$\mathcal O(\omega_0L)$ corrections.

Define the symmetric trace-free (STF) transition quadrupole
\begin{equation}
 Q_{ij}^{ab}
 \equiv
 \mathcal Q_{ij}^{ab}
 -\frac13\delta_{ij}\mathcal Q_{kk}^{ab}\,,
 \qquad
 Q_{ij}^{12}=(Q_{ij}^{21})^*\,.
 \label{eq:sm_stf_quadrupole}
\end{equation}
Since $\Lambda_{ij,kl}\delta_{kl}=0$, the on-shell vertices in
Eq.~\eqref{eq:sm_tt_cut} reduce to their STF parts. The required
integrals are
\begin{align}
 &\int d\Omega_{\mathbf k}\,
 \Lambda_{ij,kl}
 =
 \frac{8\pi}{5}
 \mathcal P_{ij,kl}^{\rm STF}\,,
 \nonumber\\
 &\int_0^\infty dq\,q^2
 \delta(q^2-\omega_0^2)
 =
 \frac{\omega_0}{2}\,,
 \label{eq:sm_cut_integrals}
\end{align}
where
\begin{equation}
 \mathcal P_{ij,kl}^{\rm STF}
 \equiv
 \delta_{i(k}\delta_{l)j}
 -\frac13\delta_{ij}\delta_{kl}\,.
 \label{eq:sm_stf_projector}
\end{equation}
It follows that
\begin{equation}
 \operatorname{Im}\Sigma_{{\rm TT},12}
 =
 -\frac{2G}{5}\mu M_c\omega_0^5
 Q_{ij}^{12}Q_{ij}^{21}\,.
 \label{eq:sm_tt_imaginary_part}
\end{equation}
Only the TT on-shell response is dissipative.  The TT principal-value
part joins the constraint response in the conservative frequency
shift.

\subsection{Two-level amplitude equations}

Parameterise the full self-energy as
\begin{equation}
 \Sigma_{12}
 \equiv-\Delta-i\Gamma_{\rm eff}\,,
 \label{eq:sm_shift_and_rate}
\end{equation}
where $\Delta$ and $\Gamma_{\rm eff}$ are the real shift and collective rate used in the main text. Defining $|Q^{21}|^2\equiv Q_{ij}^{12}Q_{ij}^{21}$, Eqs.~\eqref{eq:sm_constraint_no_imaginary} and \eqref{eq:sm_tt_imaginary_part} give
\begin{equation}
 \Gamma_{\rm eff}
 =
 -\operatorname{Im}\Sigma_{12}
 =
 \frac{2G}{5}\mu M_c\omega_0^5|Q^{21}|^2\,.
 \label{eq:sm_collective_rate}
\end{equation}
For an incident tidal field
$E_{ij}^{\mathrm{seed},(+)}(t)= \omega_{\mathrm s}^2 h_E(t)e_{ij}^{(\lambda)} e^{-i\omega_{\mathrm s}t}/2$, the tidal interaction gives
\begin{align}
 V_{12}^{(+),{\rm seed}}(t)
 &=\frac12\Omega_R(t)e^{-i\omega_{\rm s}t}\,,\nonumber\\
 \Omega_R(t)&=\frac{\mu\omega_{\mathrm s}^2}{2}Q_{ij}^{12}e_{ij}^{(\lambda)}h_E(t)\,.
 \label{eq:sm_seed_vertex}
\end{align}

Combining this drive with the self-field under the rotating-wave approximation of Eq.~\eqref{eq:sm_two_level_equations}, and using $V_{21}^{(-)}=(V_{12}^{(+)})^*$ and $\delta=\omega_{\rm s}-\omega_0$, we obtain
\begin{align}
 i\dot c_1
 &=\frac12\Big[\Omega_R(t)e^{-i\delta t}
 -(\Delta+i\Gamma_{\rm eff})c_1c_2^*\Big]c_2\,,
 \nonumber\\
 i\dot c_2
 &=\frac12\Big[\Omega_R^*(t)e^{i\delta t}
 -(\Delta-i\Gamma_{\rm eff})c_1^*c_2\Big]c_1\,.
 \label{eq:sm_full_amplitude_equations}
\end{align}
The generally complex Rabi amplitude $\Omega_R$ drives coherent population transfer, while $\Delta$ shifts the coherence phase. The imaginary part of the self-energy produces the autonomous gain.

Notice that as in the main text, we omit the diagonal matrix elements $V_{aa}$. Write the diagonal matrix elements as
$V_{aa}(t)=\overline V_{aa}+V_{aa}^{\rm osc}(t)$. The constant parts are absorbed into the level energies, $\omega_a\to\omega_a+\overline V_{aa}$, while the nonresonant oscillating parts are discarded at leading rotating-wave order.

\section{Population dynamics and seed-induced ignition}
\label{sec:sm_full_dynamics}

Neglecting single-level BH growth and absorption, differentiating $n_a=|c_a|^2$ using Eq.~\eqref{eq:sm_full_amplitude_equations} gives
\begin{align}
 \dot n_1
 &=-\Gamma_{\rm eff}n_1n_2
 -\operatorname{Im}\!\left[\Omega_R^*(t)e^{i\delta t}\zeta\right]\,,
 \nonumber\\
 \dot n_2
 &=\Gamma_{\rm eff}n_1n_2
 +\operatorname{Im}\!\left[\Omega_R^*(t)e^{i\delta t}\zeta\right]\,.
 \label{eq:sm_driven_population_coherence}
\end{align}
These equations conserve $n_1+n_2=1$, with $\zeta=c_2^*c_1$. The real self-energy cancels from the direct population derivatives but affects the relative phase that controls the seed-driven transfer. While the seed is present, this phase is determined by the amplitude equations~\eqref{eq:sm_full_amplitude_equations}.

The seed term fixes the initial occupation $n_2^{\rm seed}$ and
phase; once the seed has passed, $\Omega_R=0$, and Eq.~\eqref{eq:sm_driven_population_coherence} reduces to
\begin{equation}
 \dot n_1=-\Gamma_{\rm eff}n_1n_2\,,
 \qquad
 \dot n_2=\Gamma_{\rm eff}n_1n_2\,.
 \label{eq:sm_population_equations}
\end{equation}
Neither the detuning nor the conservative shift enters these
equations. Let the seed end at $t=t_{\rm s}$ and write
$n_2^{\rm seed}=n_2(t_{\rm s})$.  For $t\geq t_{\rm s}$,
\begin{equation}
 n_2(t)=\frac{n_2^{\rm seed}e^{\Gamma_{\rm eff}(t-t_{\rm s})}}
 {1-n_2^{\rm seed}+n_2^{\rm seed}e^{\Gamma_{\rm eff}(t-t_{\rm s})}}\,.
 \label{eq:sm_post_seed_solution}
\end{equation}
Defining $t_p=\Gamma_{\rm eff}^{-1}$ and $n_2(t_D)=1/2$ gives
\begin{equation}
 t_D=t_{\rm s}+t_p\ln\frac{1-n_2^{\rm seed}}{n_2^{\rm seed}}\,.
 \label{eq:sm_ignition_delay}
\end{equation}
The populations and coherence magnitude then take the universal form
\begin{align}
 n_1(t)&=\frac12\left[1-\tanh\frac{t-t_D}{2t_p}\right]\,,
 \nonumber\\
 n_2(t)&=\frac12\left[1+\tanh\frac{t-t_D}{2t_p}\right]\,,
 \nonumber\\
 |\zeta(t)|&=\frac12\operatorname{sech}\frac{t-t_D}{2t_p}\,.
 \label{eq:sm_logistic_coherence}
\end{align}
Writing $\zeta=|\zeta|e^{i\phi}$, the same amplitude equations give
\begin{align}
 \frac{\dot\zeta}{\zeta}
 &=\frac{\Gamma_{\rm eff}-i\Delta}{2}(n_1-n_2)\,,
 \nonumber\\
 \dot\phi&=\frac{\Delta}{2}(n_2-n_1)\,.
 \label{eq:sm_coherence_phase_evolution}
\end{align}
For constant $\Delta$, substituting Eq.~\eqref{eq:sm_logistic_coherence}
and defining $\phi_D\equiv\phi(t_D)$ yields the full complex coherence
\begin{equation}
 \begin{aligned}
 \zeta(t)&=\frac12\operatorname{sech}x
 \exp\!\left[
 i\phi_D+i\frac{\Delta}{\Gamma_{\rm eff}}\ln\cosh x
 \right]\,,\\
 x&\equiv\frac{t-t_D}{2t_p}\,.
 \end{aligned}
 \label{eq:sm_complex_coherence}
\end{equation}
For $n_2^{\rm seed}\ll1$, $t_D-t_{\rm s}\simeq-t_p\ln n_2^{\rm seed}$.

The ignition stage depends on the Rabi envelope $\Omega_R(t)$,
the detuning $\delta$, and the conservative shift $\Delta$.
The transformation $c_1=a_1e^{-i\delta t/2}$ and
$c_2=a_2e^{i\delta t/2}$ removes the explicit carrier phases:
\begin{align}
 \dot a_1
 &=\frac{i\delta}{2}a_1-\frac{i}{2}\Omega_R(t)a_2
 +\frac{i\Delta-\Gamma_{\rm eff}}{2}|a_2|^2a_1\,,
 \nonumber\\
 \dot a_2
 &=-\frac{i\delta}{2}a_2-\frac{i}{2}\Omega_R^*(t)a_1
 +\frac{i\Delta+\Gamma_{\rm eff}}{2}|a_1|^2a_2\,.
 \label{eq:sm_rotating_amplitudes}
\end{align}
Consider a weak seed acting from $t_{\rm i}$ to $t_{\rm s}$, with
$\Delta t_{\rm s}=t_{\rm s}-t_{\rm i}$ and
$\Gamma_{\rm eff}\Delta t_{\rm s}\ll1$. For an initially inverted
cloud, $|a_1|^2\simeq1$ and $|a_2|^2\ll1$, growth during preparation
can be neglected, giving
\begin{align}
 \dot a_1&\simeq\frac{i\delta}{2}a_1\,,\nonumber\\
 \dot a_2&\simeq\frac{i(\Delta-\delta)}{2}a_2
 -\frac{i}{2}\Omega_R^*(t)a_1\,.
 \label{eq:sm_linear_seed_amplitudes}
\end{align}
Both levels accumulate phase. Their relative phase in the drive
therefore selects the effective detuning
\begin{equation}
 \delta_{\rm eff}
 \equiv\frac{\delta}{2}-\frac{\Delta-\delta}{2}
 =\delta-\frac{\Delta}{2}\,.
 \label{eq:sm_ignition_detuning}
\end{equation}
For approximately constant $\delta_{\rm eff}$ and zero initial
lower-level amplitude, integrating
Eq.~\eqref{eq:sm_linear_seed_amplitudes} gives
\begin{equation}
 n_2^{\rm seed}=|a_2(t_{\rm s})|^2
 \simeq\frac14\left|
 \int_{t_{\rm i}}^{t_{\rm s}}dt\,
 \Omega_R(t)e^{-i\delta_{\rm eff}t}
 \right|^2\,,
 \label{eq:sm_ignition_seed_integral}
\end{equation}
where we have taken the complex conjugate inside the modulus. Thus the prepared population samples the Fourier component of the
Rabi envelope at the shifted resonance.

For a rectangular seed of constant complex amplitude $\Omega_0$
and duration $\Delta t_{\rm s}$,
\begin{equation}
 n_2^{\rm seed}\simeq
 \frac{|\Omega_0|^2\Delta t_{\rm s}^2}{4}
 \operatorname{sinc}^2\!\left(\frac{\delta_{\rm eff}\Delta t_{\rm s}}{2}\right)\,,
 \label{eq:sm_ignition_square_seed}
\end{equation}
where $\operatorname{sinc}x=\sin x/x$.  On shifted resonance,
$\delta_{\rm eff}=0$, Eq.~\eqref{eq:sm_ignition_delay} gives
\begin{equation}
 \frac{t_D-t_{\rm s}}{t_p}
 \simeq2\ln\frac{2}{|\Omega_0|\Delta t_{\rm s}}\,.
 \label{eq:sm_ignition_resonant_delay}
\end{equation}
For a Gaussian envelope
$\Omega_R(t)=\Omega_0\exp[-(t-t_0)^2/(2s_t^2)]$, centred at $t_0$
with width $s_t\ll t_p$, integration over the pulse gives
\begin{equation}
 n_2^{\rm seed}\simeq\frac{\pi}{2}|\Omega_0|^2s_t^2
 e^{-\delta_{\rm eff}^2s_t^2}\,.
 \label{eq:sm_ignition_gaussian_seed}
\end{equation}
The conservative shift affects the preparation efficiency and hence the delay through Eq.~\eqref{eq:sm_ignition_delay}.
After the drive vanishes, it still changes the coherence phase, but the population profile is the logistic solution above.

\section{Energy-balance check}
\label{sec:sm_energy_balance}

The cloud energy is
\begin{equation}
 E_{\rm cloud}
 =
 N(\omega_1n_1+\omega_2n_2)\,.
 \label{eq:sm_cloud_energy}
\end{equation}
With $N=M_c/\mu$, Eq.~\eqref{eq:sm_population_equations} gives
\begin{equation}
 -\dot E_{\rm cloud}
 =N\omega_0\dot n_2
 =N\omega_0\Gamma_{\rm eff}|\zeta|^2\,,
 \label{eq:sm_cloud_energy_loss}
\end{equation}
where $|\zeta|^2=n_1n_2$.

We independently compute the radiated power from the polarizations in Eq.~\eqref{eq:main_pulse_waveform}. Their angular factors obey
\begin{equation}
 \int \mathrm{d}\Omega\left[
 \left(\frac{1+\cos^2\iota}{2}\right)^2+\cos^2\iota
 \right]=\frac{16\pi}{5}\,.
 \label{eq:sm_flux_angular_integral}
\end{equation}
Using $\operatorname{sech}[(t_{\rm r}-t_D)/(2t_p)]=2|\zeta(t_{\rm r})|$
and the amplitude in Eq.~\eqref{eq:main_strain_amplitude}, and neglecting derivatives of the slow envelope, we obtain
\begin{align}
 P_{\rm GW}(t_{\rm r})
 &=\frac{d^2}{16\pi G}\int d\Omega\,
 \left\langle\dot h_+^2+\dot h_\times^2\right\rangle_{\rm cyc}
 \nonumber\\
 &=\frac{8G}{5}M_c^2\omega_0^6
 |Q_{xx}^{21}|^2|\zeta(t_{\rm r})|^2\,.
 \label{eq:sm_waveform_power}
\end{align}
Here the brackets denote an average over the carrier cycle, and the power is labelled by the retarded source time.  For the two $\Delta m=-2$ channels in Table~\ref{tab:channel_coefficients}, $Q_{ij}^{12}Q_{ij}^{21}=4|Q_{xx}^{21}|^2$, so Eq.~\eqref{eq:sm_collective_rate} yields
\begin{equation}
 \Gamma_{\rm eff}=\frac{8G}{5}\mu M_c\omega_0^5|Q_{xx}^{21}|^2\,.
 \label{eq:sm_channel_balance_rate}
\end{equation}
Thus, substituting Eq.~\eqref{eq:sm_channel_balance_rate} into Eq.~\eqref{eq:sm_cloud_energy_loss}, we obtain
\begin{equation}
 P_{\rm GW}=N\omega_0\Gamma_{\rm eff}|\zeta|^2
 =-\dot E_{\rm cloud}\,.
 \label{eq:sm_energy_balance}
\end{equation}
This equality verifies the consistency between the collective transfer rate and the outgoing GW power; a complete population transfer consequently radiates the energy $N\omega_0$.


\section{Details of parameter scan}
\label{sec:g_cloud_mass}

We estimate the available $544$ cloud mass from sequential
superradiant saturation, following Ref.~\cite{Kim:2025wwj}.
This estimate neglects cloud depletion during growth and
reabsorption of earlier clouds.  Let $M^{(3)}$ and
$\tilde a^{(3)}$ denote the BH mass and spin after the
$m=3$ stage, and let $M$ and $\tilde a^{(4)}$ denote them after
the $m=4$ stage.  Here $M_c$ is the newly accumulated $544$ cloud
mass, and $\sigma=M_c/M$ is normalised to the final BH mass.
Mass conservation gives $M^{(3)}=M+M_c=M(1+\sigma)$, and accordingly, 
\begin{equation}
 \alpha^{(3)}=GM^{(3)}\mu=\alpha(1+\sigma)\,.
 \label{eq:g_mass_balance}
\end{equation}
At leading order, saturation occurs at $m\Omega_H=\mu$.  Therefore
\begin{align}
 \tilde a^{(3)}
 &=\frac{4\alpha^{(3)}/3}{1+4(\alpha^{(3)}/3)^2}\,,\nonumber\\
 \tilde a^{(4)}
 &=\frac{\alpha}{1+\alpha^2/4}\,.
 \label{eq:g_saturation_spins}
\end{align}
Each extracted $544$ quantum carries energy $\mu$ and angular
momentum $m=4$.  With $J^{(k)}$ the BH angular momentum
at stage $k$, angular-momentum conservation reads
\begin{equation}
 J^{(3)}-J^{(4)}=\frac{4M_c}{\mu}\,.
 \label{eq:g_angular_momentum_balance}
\end{equation}
Using $J^{(3)}=G[M^{(3)}]^2\tilde a^{(3)}$ and
$J^{(4)}=GM^2\tilde a^{(4)}$ reduces this to
\begin{equation}
 \tilde a^{(3)}(1+\sigma)^2-\tilde a^{(4)}
 -\frac{4\sigma}{\alpha}=0\,.
 \label{eq:g_sigma_equation}
\end{equation}
For fixed final $\alpha$, the only unknown in Eq.~\eqref{eq:g_sigma_equation} is $\sigma$: the mass balance \eqref{eq:g_mass_balance} gives $\alpha^{(3)}=\alpha(1+\sigma)$, which fixes $\tilde a^{(3)}$ through \eqref{eq:g_saturation_spins}, whereas $\tilde a^{(4)}$ is determined by $\alpha$ alone.  Solving \eqref{eq:g_sigma_equation} then gives $\sigma(\alpha)$.

The single-level rates
$\gamma_{n\ell m}=\operatorname{Im}\omega_{n\ell m}^{\Psi}$ follow from Eq.~\eqref{eq:sm_superradiant_rate}. The $544$ cloud forms on the preceding $m=3$ background, of mass $M^{(3)}=M(1+\sigma)$, whereas the $322$ level is absorbed on the final background, so the parameter scan uses
\begin{align}
 t_{\rm SR}
 &=\frac{1}{2\gamma_{544}(\alpha^{(3)},\tilde a^{(3)})}\,,
 \nonumber\\
 \Gamma_2
 &=-2\gamma_{322}(\alpha,\tilde a^{(4)})\,.
 \label{eq:g_growth_absorption}
\end{align}
Here $t_{\rm SR}$ is the occupation-number e-folding time. At fixed growth rate, the formation timescale is
\begin{equation}
 t_{\rm form}
 =t_{\rm SR}\ln\!\left(\frac{N_{\rm sat}}{N_{\rm seed}}\right)
 \simeq100t_{\rm SR}\,,
 \label{eq:g_formation_time}
\end{equation}
where $N_{\rm seed}$ is the initial cloud occupation and $N_{\rm sat}=M_c/\mu=\sigma M/\mu$ is its saturated value. We adopt $\mathcal O(100)$ occupation-number e-folds as a representative estimate~\cite{Fernandez:2019qbj,Siemonsen:2022darkphoton}. The shaded region of Fig.~\ref{fig:parameter_space} requires
$t_{\rm form}\le10^{10}\,{\rm yr}$ and
$\Gamma_{\rm eff}>\Gamma_2$.

\section{Far-zone waveform and sensitivity estimates}
\label{sec:far_zone}

\subsection{Quadrupole projection and polarizations}
\label{sec:g_waveform}

At leading nonrelativistic order, the mass quadrupole
$M_{ij}=\int d^3\mathbf x\,\rho x_ix_j$, with
$\rho\simeq\mu|\Psi|^2$, contains the beat component
\begin{equation}
 M_{ij}(t)=2M_c\,\operatorname{Re}
 \left[\zeta(t)\mathcal Q_{ij}^{21}e^{-i\omega_0t}\right]\,.
 \label{eq:g_beat_quadrupole}
\end{equation}
The radiation in direction $\hat{\mathbf n}$ follows from Eq.~\eqref{eq:main_far_zone_projection}. The projector defined there satisfies $\Lambda_{ij,kl}\delta_{kl}=0$, so that $\Lambda_{ij,kl}\mathcal Q_{kl}^{21}=\Lambda_{ij,kl}Q_{kl}^{21}$.
Keeping the carrier derivatives gives
\begin{align}
 h_{ij}^{\rm TT}(t,d\hat{\mathbf n})
 \simeq{}&-\frac{4GM_c\omega_0^2}{d}
 \operatorname{Re}\Big[
 \zeta(t_{\rm r})\Lambda_{ij,kl}(\hat{\mathbf n})
 \nonumber\\
 &\hspace{18mm}\times Q_{kl}^{21}e^{-i\omega_0t_{\rm r}}
 \Big]\,,
 \label{eq:g_projected_wave}
\end{align}
where $t_{\rm r}=t-d$ and the coherence varies slowly compared with $\omega_0^{-1}$.

For $\Delta m=m_2-m_1=-2$, take the spin axis along $z$ and use $Y_{\ell m}\propto e^{im\varphi}$. The overlap $\psi_2^*\psi_1\propto e^{2i\varphi}$ fixes the Cartesian tensor to
\begin{equation}
 Q_{ij}^{21}=Q_{xx}^{21}
 \begin{pmatrix}
 1&i&0\\
 i&-1&0\\
 0&0&0
 \end{pmatrix}\,.
 \label{eq:g_quadrupole_tensor}
\end{equation}
Choose the observer at azimuth zero and define
\begin{align}
 \hat{\mathbf n}&=(\sin\iota,0,\cos\iota)\,,\nonumber\\
 \mathbf e_\theta&=(\cos\iota,0,-\sin\iota)\,,\qquad
 \mathbf e_\varphi=(0,1,0)\,.
 \label{eq:g_polarization_basis}
\end{align}
The polarization tensors are
$e^+_{ij}=e_{\theta i}e_{\theta j}-e_{\varphi i}e_{\varphi j}$
and
$e^\times_{ij}=e_{\theta i}e_{\varphi j}+e_{\varphi i}e_{\theta j}$,
with $h_A=e^A_{ij}h_{ij}^{\rm TT}/2$ for $A=+,\times$.
They are already transverse and traceless, and
\begin{align}
 e^+_{ij}Q_{ij}^{21}&=(1+\cos^2\iota)Q_{xx}^{21}\,,\nonumber\\
 e^\times_{ij}Q_{ij}^{21}&=2i\cos\iota\,Q_{xx}^{21}\,.
 \label{eq:g_polarization_contractions}
\end{align}
Thus, with $h_0$ defined in Eq.~\eqref{eq:main_strain_amplitude}, write $\zeta=|\zeta|e^{i\phi}$ and define
$\vartheta(t)\equiv\omega_0t-\phi(t)$.  The polarizations are
\begin{align}
 h_+(t)&\simeq-h_0(1+\cos^2\iota)
 |\zeta(t_{\rm r})|\cos\vartheta(t_{\rm r})\,,\nonumber\\
 h_\times(t)&\simeq-2h_0\cos\iota
 |\zeta(t_{\rm r})|\sin\vartheta(t_{\rm r})\,.
 \label{eq:g_polarizations}
\end{align}

Using Eq.~\eqref{eq:sm_complex_coherence}, the carrier phase after the seed is
\begin{align}
 \vartheta(t_{\rm r})
 &=\omega_0t_{\rm r}-\phi_D
 -\frac{\Delta}{\Gamma_{\rm eff}}
 \ln\cosh\!\left(\frac{t_{\rm r}-t_D}{2t_p}\right)\,,
 \nonumber\\
 \dot\vartheta(t_{\rm r})
 &=\omega_0-\frac{\Delta}{2}
 \tanh\!\left(\frac{t_{\rm r}-t_D}{2t_p}\right)\,.
 \label{eq:g_carrier_phase}
\end{align}
The representative waveform and spectrum of Sec.~IV omit this phase
modulation while retaining the universal coherence envelope.

For $|1\rangle=|544\rangle$ and $|2\rangle=|322\rangle$, the angular
integral in Eq.~\eqref{eq:g_quadrupole_tensor} gives
\begin{align}
 Q_{xx}^{21}
 &=\frac{1}{\sqrt{21}}\int_0^\infty dr\,r^4R_{32}(r)R_{54}(r)
 \nonumber\\
 &=\frac{102515625}{4194304}r_B^2
 \simeq24.4\,r_B^2\,,
 \label{eq:g_channel_quadrupole}
\end{align}
using the normalised radial functions in
Eq.~\eqref{eq:sm_radial_mode}.  The radial functions are chosen real
with their standard hydrogenic phases.
The tensor norm is
$Q_{ij}^{12}Q_{ij}^{21}=4|Q_{xx}^{21}|^2$.
With $\omega_0=(8/225)\mu\alpha^2$, this yields
\begin{align}
 \Gamma_{\rm eff}
 &\simeq5.4\times10^{-5}\,
 \frac{\sigma\alpha^8}{GM}\,,\nonumber\\
 h_0&\simeq0.06\,\sigma\alpha^2\frac{GM}{d}\,.
 \label{eq:g_channel_rate_strain}
\end{align}
These expressions give the normalisation used in
Eq.~\eqref{eq:main_h0_benchmark}.

\subsection{Pulse spectrum and high-frequency sensitivity}
\label{sec:g_detectability}

Let $F_+$ and $F_\times$ be the antenna factors. The detector strain $h=F_+h_++F_\times h_\times$ has amplitude $h_{\rm eff}={\cal A}_{\rm det}h_0$, with
\begin{equation}
 {\cal A}_{\rm det}^2
 =F_+^2\frac{(1+\cos^2\iota)^2}{4}
 +F_\times^2\cos^2\iota\,.
 \label{eq:g_detector_factor}
\end{equation}
With the retarded time measured from the pulse peak, we write the detector strain as
\begin{equation}
 h(t)=h_{\rm eff}{\cal E}(t)
 \cos\!\left[\vartheta(t)+\phi_{\rm det}\right]\,,
 \label{eq:g_detector_strain}
\end{equation}
where ${\cal E}(t)=\operatorname{sech}\left({t}/{(2t_p)}\right)$,
$f_0=\omega_0/(2\pi)$, and $\phi_{\rm det}$ is the constant phase set by
the detector response and source orientation.

For the phase-unmodulated representative waveform,
$\vartheta(t)+\phi_{\rm det}=2\pi f_0t+\phi_0$.  Using
$\tilde h(f)=\int dt\,h(t)e^{-2\pi ift}$, for $f_0t_p\gg1$, we have
\begin{equation}
 |\tilde h(f)|\simeq
 \pi t_ph_{\rm eff}
 \operatorname{sech}\left[2\pi^2t_p(f-f_0)\right]\,.
 \label{eq:g_pulse_spectrum}
\end{equation}
The FWHM of the time-domain amplitude envelope, which is independent of the
carrier-phase modulation, is given below together with the power FWHM of the
phase-unmodulated spectrum in Eq.~\eqref{eq:g_pulse_spectrum}:
\begin{align}
 \Delta t_{\rm FWHM}&=4t_p\operatorname{arcosh}(2)\,,\nonumber\\
 \Delta f_{\rm FWHM}
 &=\frac{\operatorname{arcosh}(\sqrt2)}{\pi^2t_p}\,.
 \label{eq:g_pulse_widths}
\end{align}
At the benchmark, the envelope width is $194\,{\rm s}$, while the
phase-unmodulated spectrum has width $2.42\,{\rm mHz}$, as shown in
Fig.~\ref{fig:waveform}.

For a one-sided detector noise power spectral density $S_n(f)$, the optimal matched-filter norm is
\begin{equation}
 {\rm SNR}^2=4\int_0^\infty\frac{|\tilde h(f)|^2}{S_n(f)}\,df\,.
 \label{eq:g_matched_filter}
\end{equation}
If the noise is approximately constant across the signal bandwidth---given by
$\Delta f_{\rm FWHM}$ for the phase-unmodulated waveform---then
$S_n(f)\simeq S_n(f_0)$. We can approximate
\begin{align}
 {\rm SNR}^2
 &\simeq\frac{2}{S_n(f_0)}\int_{\rm obs}dt\,h^2(t)
 \nonumber\\
 &\simeq\frac{h_{\rm eff}^2}{S_n(f_0)}
 \int_{\rm obs}dt\,{\cal E}^2(t)\,.
 \label{eq:g_snr_time_domain}
\end{align}
For a window of duration $T_{\rm obs}$ centred on the peak,
\begin{equation}
 \int_{-T_{\rm obs}/2}^{T_{\rm obs}/2}dt\,{\cal E}^2(t)
 =4t_p\tanh\left(\frac{T_{\rm obs}}{4t_p}\right)\,,
 \label{eq:g_window_integral}
\end{equation}
which yields Eq.~\eqref{eq:main_pulse_snr}.



	\bibliographystyle{utphys}
	\bibliography{ref}

@article{Bian:2025gwcosmology,
  author = {Bian, Ligong and Cai, Rong-Gen and Dong, Yu-Qi and Gao, Qing and Gong, Yungui and Guo, Zong-Kuan and Huang, Qing-Guo and Kang, Zhaofeng and Li, Li and Liu, Jing and Liu, Lang and Liu, Yu-Xiao and Lu, Xuchen and Peng, Zhi-Zhang and Qiao, Jin and Wu, Puxun and Wu, Yue-Liang and Yu, Jiang-Hao and Yuan, Chen and Zhang, Chao and Zhu, Tao},
  title = {Gravitational wave cosmology},
  journal = {Sci. China Phys. Mech. Astron.},
  volume = {69},
  pages = {210401},
  year = {2026},
  doi = {10.1007/s11433-025-2740-8}
}

@article{PhysRev.93.99,
  title = {Coherence in Spontaneous Radiation Processes},
  author = {Dicke, R. H.},
  journal = {Phys. Rev.},
  volume = {93},
  issue = {1},
  pages = {99--110},
  numpages = {0},
  year = {1954},
  month = {Jan},
  publisher = {American Physical Society},
  doi = {10.1103/PhysRev.93.99},
}

@article{PhysRevA.14.1169,
  title = {Theory of superradiance in an extended, optically thick medium},
  author = {MacGillivray, J. C. and Feld, M. S.},
  journal = {Phys. Rev. A},
  volume = {14},
  issue = {3},
  pages = {1169--1189},
  numpages = {0},
  year = {1976},
  month = {Sep},
  publisher = {American Physical Society},
  doi = {10.1103/PhysRevA.14.1169},
}

@article{PhysRevA.11.1507,
  title = {Cooperative radiation processes in two-level systems: Superfluorescence},
  author = {Bonifacio, R. and Lugiato, L. A.},
  journal = {Phys. Rev. A},
  volume = {11},
  issue = {5},
  pages = {1507--1521},
  numpages = {0},
  year = {1975},
  month = {May},
  publisher = {American Physical Society},
  doi = {10.1103/PhysRevA.11.1507},
}

@article{GROSS1982301,
    title = {Superradiance: An essay on the theory of collective spontaneous emission},
    journal = {Physics Reports},
    volume = {93},
    number = {5},
    pages = {301-396},
    year = {1982},
    issn = {0370-1573},
    doi = {https://doi.org/10.1016/0370-1573(82)90102-8},
    author = {M. Gross and S. Haroche}
}

@article{Zeldovich:1971ffh,
    author = "Zeldovich, Yakov Borisovich",
    title = "{Generation of Waves by a Rotating Body}",
    journal = "Soviet Journal of Experimental and Theoretical Physics Letters",
    volume = "14",
    pages = "180",
    year = "1971"
}

@article{Starobinskii:1973vzb,
    author = "Starobinskii, A. A.",
    title = "{Amplification of waves during reflection from a rotating ''black hole''}",
    journal = "Sov. Phys. JETP",
    volume = "37",
    number = "1",
    pages = "28--32",
    year = "1973"
}

@article{Press:1972zz,
    author = "Press, William H. and Teukolsky, Saul A.",
    title = "{Floating Orbits, Superradiant Scattering and the Black-hole Bomb}",
    doi = "10.1038/238211a0",
    journal = "Nature",
    volume = "238",
    pages = "211--212",
    year = "1972"
}

@article{Brito:2015oca,
    author = "Brito, Richard and Cardoso, Vitor and Pani, Paolo",
    title = "{Superradiance}: {New Frontiers in Black Hole
Physics}",
    eprint = "1501.06570",
    archivePrefix = "arXiv",
    primaryClass = "gr-qc",
    doi = "10.1007/978-3-319-19000-6",
    journal = "Lect. Notes Phys.",
    volume = "906",
    pages = "pp.1--237",
    year = "2015"
}

@article{Arvanitaki:2014wva,
    author = "Arvanitaki, Asimina and Baryakhtar, Masha and Huang, Xinlu",
    title = "{Discovering the QCD Axion with Black Holes and Gravitational Waves}",
    eprint = "1411.2263",
    archivePrefix = "arXiv",
    primaryClass = "hep-ph",
    doi = "10.1103/PhysRevD.91.084011",
    journal = "Phys. Rev. D",
    volume = "91",
    number = "8",
    pages = "084011",
    year = "2015"
}

@article{Arvanitaki:2009fg,
    author = "Arvanitaki, Asimina and Dimopoulos, Savas and Dubovsky, Sergei and Kaloper, Nemanja and March-Russell, John",
    title = "{String Axiverse}",
    eprint = "0905.4720",
    archivePrefix = "arXiv",
    primaryClass = "hep-th",
    doi = "10.1103/PhysRevD.81.123530",
    journal = "Phys. Rev. D",
    volume = "81",
    pages = "123530",
    year = "2010"
}

@article{Arvanitaki:2010sy,
    author = "Arvanitaki, Asimina and Dubovsky, Sergei",
    title = "{Exploring the String Axiverse with Precision Black Hole Physics}",
    eprint = "1004.3558",
    archivePrefix = "arXiv",
    primaryClass = "hep-th",
    doi = "10.1103/PhysRevD.83.044026",
    journal = "Phys. Rev. D",
    volume = "83",
    pages = "044026",
    year = "2011"
}

@article{Baumann:2019eav,
  title = {The {{Spectra}} of {{Gravitational Atoms}}},
  author = {Baumann, Daniel and Chia, Horng Sheng and Stout, John and {ter Haar}, Lotte},
  year = {2019},
  month = dec,
  journal = {JCAP},
  volume = {12},
  eprint = {1908.10370},
  primaryclass = {gr-qc},
  pages = {006},
  doi = {10.1088/1475-7516/2019/12/006},
  archiveprefix = {arXiv}
}

@article{Fernandez:2019qbj,
  title = {Superradiance and the {{Spins}} of {{Black Holes}} from {{LIGO}} and {{X-ray}} Binaries},
  author = {Fernandez, Nicolas and Ghalsasi, Akshay and Profumo, Stefano},
  year = {2019},
  month = nov,
  eprint = {1911.07862},
  primaryclass = {hep-ph},
  doi = {10.48550/arXiv.1911.07862},
  archiveprefix = {arXiv}
}

@article{Siemonsen:2022darkphoton,
  author = {Siemonsen, Nils and Mondino, Cristina and Ega{\~n}a-Ugrinovic, Daniel and Huang, Junwu and Baryakhtar, Masha and East, William E.},
  title = {Dark photon superradiance: Electrodynamics and multimessenger signals},
  eprint = {2212.09772},
  archiveprefix = {arXiv},
  primaryclass = {astro-ph.HE},
  doi = {10.1103/PhysRevD.107.075025},
  journal = {Phys. Rev. D},
  volume = {107},
  pages = {075025},
  year = {2023}
}

@article{Stott:2018opm,
  title = {Black Hole Spin Constraints on the Mass Spectrum and Number of Axionlike Fields},
  author = {Stott, Matthew J. and Marsh, David J. E.},
  year = {2018},
  month = oct,
  journal = {Phys. Rev. D},
  volume = {98},
  number = {8},
  eprint = {1805.02016},
  primaryclass = {hep-ph},
  pages = {083006},
  issn = {2470-0010, 2470-0029},
  doi = {10.1103/PhysRevD.98.083006},
  archiveprefix = {arXiv}
}

@article{Cheng:2022jsw,
    author = "Cheng, Lei-dong and Zhang, Hong and Bao, Shou-shan",
    title = "{Constraints on an axionlike particle from black hole spin superradiance}",
    eprint = "2201.11338",
    archivePrefix = "arXiv",
    primaryClass = "gr-qc",
    doi = "10.1103/PhysRevD.107.063021",
    journal = "Phys. Rev. D",
    volume = "107",
    number = "6",
    pages = "063021",
    year = "2023"
}

@article{Davoudiasl:2019nlo,
    author = "Davoudiasl, Hooman and Denton, Peter B",
    title = "{Ultralight Boson Dark Matter and Event Horizon Telescope Observations of M87*}",
    eprint = "1904.09242",
    archivePrefix = "arXiv",
    primaryClass = "astro-ph.CO",
    doi = "10.1103/PhysRevLett.123.021102",
    journal = "Phys. Rev. Lett.",
    volume = "123",
    number = "2",
    pages = "021102",
    year = "2019"
}

@article{Saha:2022hcd,
    author = "Saha, Akash Kumar and Parashari, Priyank and Maity, Tarak Nath and Dubey, Abhishek and Bouri, Subhadip and Laha, Ranjan",
    title = "{Bounds on ultralight bosons from the Event Horizon Telescope observation of Sgr A$^*$}",
    eprint = "2208.03530",
    archivePrefix = "arXiv",
    primaryClass = "astro-ph.HE",
    doi = "10.1140/epjc/s10052-024-13239-x",
    journal = "Eur. Phys. J. C",
    volume = "84",
    number = "9",
    pages = "901",
    year = "2024"
}

@article{Witte:2024drg,
    author = "Witte, Samuel J. and Mummery, Andrew",
    title = "{Stepping up superradiance constraints on axions}",
    eprint = "2412.03655",
    archivePrefix = "arXiv",
    primaryClass = "hep-ph",
    doi = "10.1103/PhysRevD.111.083044",
    journal = "Phys. Rev. D",
    volume = "111",
    number = "8",
    pages = "083044",
    year = "2025"
}

@article{Hoof:2024quk,
    author = {Hoof, Sebastian and Marsh, David J E and Sisk-Reynés, Júlia and Matthews, James H and Reynolds, Christopher},
    title = {Getting More Out of Black Hole Superradiance: a Statistically Rigorous Approach to Ultralight Boson Constraints from Black Hole Spin Measurements},
    journal = {Monthly Notices of the Royal Astronomical Society},
    pages = {staf1564},
    year = {2025},
    month = {09},
    issn = {0035-8711},
    doi = {10.1093/mnras/staf1564},
}

@article{LIGOScientific:2021rnv,
    author = "Abbott, R. and others",
    collaboration = "LIGO Scientific, Virgo, KAGRA",
    title = "{All-sky search for gravitational wave emission from scalar boson clouds around spinning black holes in LIGO O3 data}",
    eprint = "2111.15507",
    archivePrefix = "arXiv",
    primaryClass = "astro-ph.HE",
    reportNumber = "P2100343",
    doi = "10.1103/PhysRevD.105.102001",
    journal = "Phys. Rev. D",
    volume = "105",
    number = "10",
    pages = "102001",
    year = "2022"
}

@article{Ng:2020ruv,
    author = "Ng, Ken K. Y. and Vitale, Salvatore and Hannuksela, Otto A. and Li, Tjonnie G. F.",
    title = "{Constraints on Ultralight Scalar Bosons within Black Hole Spin Measurements from the LIGO-Virgo GWTC-2}",
    eprint = "2011.06010",
    archivePrefix = "arXiv",
    primaryClass = "gr-qc",
    doi = "10.1103/PhysRevLett.126.151102",
    journal = "Phys. Rev. Lett.",
    volume = "126",
    number = "15",
    pages = "151102",
    year = "2021"
}

@article{Yuan:2022bem,
    author = "Yuan, Chen and Jiang, Yang and Huang, Qing-Guo",
    title = "{Constraints on an ultralight scalar boson from Advanced LIGO and Advanced Virgo{\textquoteright}s first three observing runs using the stochastic gravitational-wave background}",
    eprint = "2204.03482",
    archivePrefix = "arXiv",
    primaryClass = "astro-ph.CO",
    doi = "10.1103/PhysRevD.106.023020",
    journal = "Phys. Rev. D",
    volume = "106",
    number = "2",
    pages = "023020",
    year = "2022"
}

@article{Caputo:2025oap,
    author = "Caputo, Andrea and Franciolini, Gabriele and Witte, Samuel J.",
    title = "{Superradiance Constraints from GW231123}",
    eprint = "2507.21788",
    archivePrefix = "arXiv",
    primaryClass = "hep-ph",
    reportNumber = "CERN-TH-2025-147, DESY-25-110",
    month = "7",
    year = "2025"
}

@article{Aswathi:2025nxa,
    author = "Aswathi, P. S. and East, William E. and Siemonsen, Nils and Sun, Ling and Jones, Dana",
    title = "{Ultralight boson constraints from gravitational wave observations of spinning binary black holes}",
    eprint = "2507.20979",
    archivePrefix = "arXiv",
    primaryClass = "gr-qc",
    month = "7",
    year = "2025"
}

@article{Lyu:2025lue,
    author = "Lyu, Zhen-Hong and Cai, Rong-Gen and Guo, Zong-Kuan and He, Jian-Feng and Liu, Jing",
    title = "{Ring formation from black hole superradiance through repeated particle production on bound orbits}",
    eprint = "2507.03490",
    archivePrefix = "arXiv",
    primaryClass = "gr-qc",
    doi = "10.1103/3r41-xyj3",
    journal = "Phys. Rev. D",
    volume = "112",
    number = "10",
    pages = "104066",
    year = "2025"
}

@article{Lyu:2025nsd,
    author = "Lyu, Zhen-Hong and Cai, Rong-Gen and Wang, Shao-Jiang and Zeng, Xiang-Xi",
    title = "{Constraining interacting dark energy models with black hole superradiance}",
    eprint = "2511.16244",
    archivePrefix = "arXiv",
    primaryClass = "astro-ph.CO",
    doi = "10.1103/rmgx-rp87",
    journal = "Phys. Rev. D",
    volume = "113",
    number = "8",
    pages = "083041",
    year = "2026"
}

@article{Ghosh:2018gaw,
  title = {Follow-up Signals from Superradiant Instabilities of Black Hole Merger Remnants},
  author = {Ghosh, Shrobana and Berti, Emanuele and Brito, Richard and Richartz, Mauricio},
  year = {2019},
  month = may,
  journal = {Phys. Rev. D},
  volume = {99},
  number = {10},
  eprint = {1812.01620},
  primaryclass = {gr-qc},
  pages = {104030},
  doi = {10.1103/PhysRevD.99.104030},
  archiveprefix = {arXiv}
}

@article{Liu:2024mzw,
    author = "Liu, Jing",
    title = "{Gravitational laser: the stimulated radiation of gravitational waves from the clouds of ultralight bosons}",
    eprint = "2401.16096",
    archivePrefix = "arXiv",
    primaryClass = "gr-qc",
    doi = "10.1088/1572-9494/ae144b",
    journal = "Commun. Theor. Phys.",
    volume = "78",
    number = "3",
    pages = "035403",
    year = "2026"
}

@article{Guo:2022mpr,
    author = "Guo, Yin-da and Bao, Shou-shan and Zhang, Hong",
    title = "{Subdominant modes of the scalar superradiant instability and gravitational wave beats}",
    eprint = "2212.07186",
    archivePrefix = "arXiv",
    primaryClass = "gr-qc",
    reportNumber = "Phys. Rev. D 107, 075009 (2023)",
    doi = "10.1103/PhysRevD.107.075009",
    journal = "Phys. Rev. D",
    volume = "107",
    number = "7",
    pages = "075009",
    year = "2023"
}

@article{Brito:2017wnc,
    author = "Brito, Richard and Ghosh, Shrobana and Barausse, Enrico and Berti, Emanuele and Cardoso, Vitor and Dvorkin, Irina and Klein, Antoine and Pani, Paolo",
    title = "{Stochastic and resolvable gravitational waves from ultralight bosons}",
    eprint = "1706.05097",
    archivePrefix = "arXiv",
    primaryClass = "gr-qc",
    doi = "10.1103/PhysRevLett.119.131101",
    journal = "Phys. Rev. Lett.",
    volume = "119",
    number = "13",
    pages = "131101",
    year = "2017"
}

@article{Takahashi:2026ruw,
    author = "Takahashi, Takuya",
    title = "{Relativistic frequency shifts in gravitational waves from axion clouds}",
    eprint = "2604.21239",
    archivePrefix = "arXiv",
    primaryClass = "gr-qc",
    reportNumber = "RESCEU-13/26",
    doi = "10.1103/n7sy-gv28",
    journal = "Phys. Rev. D",
    volume = "114",
    number = "4",
    pages = "044008",
    year = "2026"
}

@article{Guo:2026yiz,
    author = "Guo, Yin-Da and Brito, Richard and Yuan, Chen",
    title = "{Stochastic gravitational-wave background from self-interacting superradiant clouds}",
    eprint = "2609.33596",
    archivePrefix = "arXiv",
    primaryClass = "gr-qc",
    month = "9",
    year = "2026"
}

@article{Baumann:2018vus,
  title = {Probing {{Ultralight Bosons}} with {{Binary Black Holes}}},
  author = {Baumann, Daniel and Chia, Horng Sheng and Porto, Rafael A.},
  year = {2019},
  month = feb,
  journal = {Phys. Rev. D},
  volume = {99},
  number = {4},
  eprint = {1804.03208},
  primaryclass = {gr-qc},
  pages = {044001},
  doi = {10.1103/PhysRevD.99.044001},
  archiveprefix = {arXiv}
}

@article{Baumann:2019ztm,
    author = "Baumann, Daniel and Chia, Horng Sheng and Porto, Rafael A. and Stout, John",
    title = "{Gravitational Collider Physics}",
    eprint = "1912.04932",
    archivePrefix = "arXiv",
    primaryClass = "gr-qc",
    reportNumber = "DESY-19-221, DESY 19-221",
    doi = "10.1103/PhysRevD.101.083019",
    journal = "Phys. Rev. D",
    volume = "101",
    number = "8",
    pages = "083019",
    year = "2020"
}

@article{Zhang:2019eid,
  title = {Dynamic {{Signatures}} of {{Black Hole Binaries}} with {{Superradiant Clouds}}},
  author = {Zhang, Jun and Yang, Huan},
  year = {2020},
  month = feb,
  journal = {Phys. Rev. D},
  volume = {101},
  number = {4},
  eprint = {1907.13582},
  primaryclass = {gr-qc},
  pages = {043020},
  doi = {10.1103/PhysRevD.101.043020},
  archiveprefix = {arXiv}
}

@article{Takahashi:2021yhy,
    author = "Takahashi, Takuya and Omiya, Hidetoshi and Tanaka, Takahiro",
    title = "{Axion cloud evaporation during inspiral of black hole binaries: The effects of backreaction and radiation}",
    eprint = "2112.05774",
    archivePrefix = "arXiv",
    primaryClass = "gr-qc",
    doi = "10.1093/ptep/ptac044",
    journal = "PTEP",
    volume = "2022",
    number = "4",
    pages = "043E01",
    year = "2022"
}

@article{Tong:2022bbl,
    author = "Tong, Xi and Wang, Yi and Zhu, Hui-Yu",
    title = "{Termination of superradiance from a binary companion}",
    eprint = "2205.10527",
    archivePrefix = "arXiv",
    primaryClass = "gr-qc",
    doi = "10.1103/PhysRevD.106.043002",
    journal = "Phys. Rev. D",
    volume = "106",
    number = "4",
    pages = "043002",
    year = "2022"
}

@article{Tomaselli:2024bdd,
    author = "Tomaselli, Giovanni Maria and Spieksma, Thomas F. M. and Bertone, Gianfranco",
    title = "{Resonant history of gravitational atoms in black hole binaries}",
    eprint = "2403.03147",
    archivePrefix = "arXiv",
    primaryClass = "gr-qc",
    doi = "10.1103/PhysRevD.110.064048",
    journal = "Phys. Rev. D",
    volume = "110",
    number = "6",
    pages = "064048",
    year = "2024"
}

@article{Tomaselli:2024dbw,
  title = {Legacy of Boson Clouds on Black Hole Binaries},
  author = {Tomaselli, Giovanni Maria and Spieksma, Thomas F. M. and Bertone, Gianfranco},
  year = {2024},
  month = sep,
  journal = {Phys. Rev. Lett.},
  volume = {133},
  number = {12},
  eprint = {2407.12908},
  primaryclass = {gr-qc},
  pages = {121402},
  doi = {10.1103/PhysRevLett.133.121402},
  archiveprefix = {arXiv}
}

@article{Cao:2023fyv,
    author = "Cao, Yan and Tang, Yong",
    title = "{Signatures of ultralight bosons in compact binary inspiral and outspiral}",
    eprint = "2307.05181",
    archivePrefix = "arXiv",
    primaryClass = "gr-qc",
    doi = "10.1103/PhysRevD.108.123017",
    journal = "Phys. Rev. D",
    volume = "108",
    number = "12",
    pages = "123017",
    year = "2023"
}

@article{Guo:2025pea,
    author = "Guo, Yuhao and Zhong, Zhen and Chen, Yifan and Cardoso, Vitor and Ikeda, Taishi and Zhou, Lihang",
    title = "{Ultralight Boson Ionization from Comparable-Mass Binary Black Holes}",
    eprint = "2509.09643",
    archivePrefix = "arXiv",
    primaryClass = "gr-qc",
    month = "9",
    year = "2025"
}

@article{Duque:2023seg,
    author = "Duque, Francisco and Macedo, Caio F. B. and Vicente, Rodrigo and Cardoso, Vitor",
    title = "{Extreme-Mass-Ratio Inspirals in Ultralight Dark Matter}",
    eprint = "2312.06767",
    archivePrefix = "arXiv",
    primaryClass = "gr-qc",
    doi = "10.1103/PhysRevLett.133.121404",
    journal = "Phys. Rev. Lett.",
    volume = "133",
    number = "12",
    pages = "121404",
    year = "2024"
}

@article{Tomaselli:2023ysb,
    author = "Tomaselli, Giovanni Maria and Spieksma, Thomas F. M. and Bertone, Gianfranco",
    title = "{Dynamical friction in gravitational atoms}",
    eprint = "2305.15460",
    archivePrefix = "arXiv",
    primaryClass = "gr-qc",
    doi = "10.1088/1475-7516/2023/07/070",
    journal = "JCAP",
    volume = "07",
    pages = "070",
    year = "2023"
}

@article{Baumann:2021fkf,
    author = "Baumann, Daniel and Bertone, Gianfranco and Stout, John and Tomaselli, Giovanni Maria",
    title = "{Ionization of gravitational atoms}",
    eprint = "2112.14777",
    archivePrefix = "arXiv",
    primaryClass = "gr-qc",
    doi = "10.1103/PhysRevD.105.115036",
    journal = "Phys. Rev. D",
    volume = "105",
    number = "11",
    pages = "115036",
    year = "2022"
}

@article{Boskovic:2024fga,
    author = "Bo{\v{s}}kovi{\'c}, Mateja and Koschnitzke, Matthias and Porto, Rafael A.",
    title = "{Signatures of Ultralight Bosons in the Orbital Eccentricity of Binary Black Holes}",
    eprint = "2403.02415",
    archivePrefix = "arXiv",
    primaryClass = "gr-qc",
    reportNumber = "DESY-24-030",
    doi = "10.1103/PhysRevLett.133.121401",
    journal = "Phys. Rev. Lett.",
    volume = "133",
    number = "12",
    pages = "121401",
    year = "2024"
}

@article{Takahashi:2024fyq,
    author = "Takahashi, Takuya and Omiya, Hidetoshi and Tanaka, Takahiro",
    title = "{Self-interacting axion clouds around rotating black holes in binary systems}",
    eprint = "2408.08349",
    archivePrefix = "arXiv",
    primaryClass = "gr-qc",
    reportNumber = "RUP-24-12",
    doi = "10.1103/PhysRevD.110.104038",
    journal = "Phys. Rev. D",
    volume = "110",
    number = "10",
    pages = "104038",
    year = "2024"
}

@article{Berti:2019wnn,
    author = "Berti, Emanuele and Brito, Richard and Macedo, Caio F. B. and Raposo, Guilherme and Rosa, Joao Luis",
    title = "{Ultralight boson cloud depletion in binary systems}",
    eprint = "1904.03131",
    archivePrefix = "arXiv",
    primaryClass = "gr-qc",
    doi = "10.1103/PhysRevD.99.104039",
    journal = "Phys. Rev. D",
    volume = "99",
    number = "10",
    pages = "104039",
    year = "2019"
}

@article{Zhang:2018kib,
    author = "Zhang, Jun and Yang, Huan",
    title = "{Gravitational floating orbits around hairy black holes}",
    eprint = "1808.02905",
    archivePrefix = "arXiv",
    primaryClass = "gr-qc",
    doi = "10.1103/PhysRevD.99.064018",
    journal = "Phys. Rev. D",
    volume = "99",
    number = "6",
    pages = "064018",
    year = "2019"
}

@article{Kang:2024trj,
  title = {The {{Stochastic Gravitational Wave Background}} from {{Primordial Gravitational Atoms}}},
  author = {Kang, Zhaofeng and Li, Tianjun and Ye, Weitao},
  year = 2024,
  month = nov,
  journal = {JCAP},
  volume = {11},
  eprint = {2407.13385},
  primaryclass = {hep-ph},
  pages = {039},
  doi = {10.1088/1475-7516/2024/11/039},
  archiveprefix = {arXiv}
}

@article{Peng:2025zca,
    author = "Peng, Si-Tong and Zhang, Jun",
    title = "{Gravitational waves from superradiant cloud level transition}",
    eprint = "2504.00728",
    archivePrefix = "arXiv",
    primaryClass = "gr-qc",
    doi = "10.1103/s69g-1wyq",
    journal = "Phys. Rev. D",
    volume = "113",
    number = "6",
    pages = "064038",
    year = "2026"
}

@article{Guo:2024iye,
    author = "Guo, Ao and Zhang, Jun and Yang, Huan",
    title = "{Superradiant clouds may be relevant for close compact object binaries}",
    eprint = "2401.15003",
    archivePrefix = "arXiv",
    primaryClass = "gr-qc",
    doi = "10.1103/PhysRevD.110.023022",
    journal = "Phys. Rev. D",
    volume = "110",
    number = "2",
    pages = "023022",
    year = "2024"
}

@article{Guo:2025ckp,
    author = "Guo, Ao and Zhang, Qi-Yan and Yang, Huan and Zhang, Jun",
    title = "{Common envelope evolution of ultralight boson clouds}",
    eprint = "2508.18738",
    archivePrefix = "arXiv",
    primaryClass = "gr-qc",
    doi = "10.1103/g79c-mqgx",
    journal = "Phys. Rev. D",
    volume = "113",
    number = "4",
    pages = "043018",
    year = "2026"
}

@article{Li:2025qyu,
    author = "Li, Ximeng and Ren, Jing and Zhang, Xi-Li",
    title = "{Probing Boson Clouds with Supermassive Black Hole Binaries}",
    eprint = "2505.02866",
    archivePrefix = "arXiv",
    primaryClass = "hep-ph",
    month = "5",
    year = "2025"
}

@article{Kyriazis:2025fis,
    author = "Kyriazis, Antonios and Yang, Fengwei",
    title = "{Gravitational waves from resonant transitions of tidally perturbed gravitational atoms}",
    eprint = "2503.18121",
    archivePrefix = "arXiv",
    primaryClass = "hep-ph",
    doi = "10.1007/JHEP11(2025)062",
    journal = "JHEP",
    volume = "11",
    pages = "062",
    year = "2025"
}

@article{Baumann:2022pkl,
    author = "Baumann, Daniel and Bertone, Gianfranco and Stout, John and Tomaselli, Giovanni Maria",
    title = "{Sharp Signals of Boson Clouds in Black Hole Binary Inspirals}",
    eprint = "2206.01212",
    archivePrefix = "arXiv",
    primaryClass = "gr-qc",
    doi = "10.1103/PhysRevLett.128.221102",
    journal = "Phys. Rev. Lett.",
    volume = "128",
    number = "22",
    pages = "221102",
    year = "2022"
}

@article{Liang:2026njr,
    author = "Liang, Yizhi and Zhu, Mian and Han, Wen-Biao and Wei, Lianfu and Wang, Peng and Tao, Jun",
    title = "{Finite Coherence in Gravitational Waves from Tidally Excited Axion Clouds}",
    eprint = "2606.28161",
    archivePrefix = "arXiv",
    primaryClass = "gr-qc",
    month = "6",
    year = "2026"
}

@article{Li:2026wds,
    author = "Li, Yi-kun and Cui, Lang",
    title = "{Relativistic Tidal Transitions of Saturated Kerr Boson Clouds}",
    eprint = "2607.14627",
    archivePrefix = "arXiv",
    primaryClass = "astro-ph.GA",
    month = "7",
    year = "2026"
}

@article{Ikeda:2018nhb,
  title = {Blasts of Light from Axions},
  author = {Ikeda, Taishi and Brito, Richard and Cardoso, Vitor},
  year = {2019},
  month = mar,
  journal = {Phys. Rev. Lett.},
  volume = {122},
  number = {8},
  eprint = {1811.04950},
  primaryclass = {gr-qc},
  pages = {081101},
  publisher = {American Physical Society},
  doi = {10.1103/PhysRevLett.122.081101},
  urldate = {2024-12-22},
}

@article{Rosa:2017ury,
    author = "Rosa, Jo{\~a}o G. and Kephart, Thomas W.",
    title = "{Stimulated Axion Decay in Superradiant Clouds around Primordial Black Holes}",
    eprint = "1709.06581",
    archivePrefix = "arXiv",
    primaryClass = "gr-qc",
    doi = "10.1103/PhysRevLett.120.231102",
    journal = "Phys. Rev. Lett.",
    volume = "120",
    number = "23",
    pages = "231102",
    year = "2018"
}

@article{Spieksma:2023vwl,
  title = {Superradiance: {{Axionic Couplings}} and {{Plasma Effects}}},
  shorttitle = {Superradiance},
  author = {Spieksma, Thomas F. M. and Cannizzaro, Enrico and Ikeda, Taishi and Cardoso, Vitor and Chen, Yifan},
  year = {2023},
  month = sep,
  journal = {Phys. Rev. D},
  volume = {108},
  number = {6},
  eprint = {2306.16447},
  primaryclass = {gr-qc},
  pages = {063013},
  doi = {10.1103/PhysRevD.108.063013},
  archiveprefix = {arXiv}
}

@article{Bai:2025yxm,
    author = "Bai, Zhaoyu and Cardoso, Vitor and Chen, Yifan and Do, Tuan and Hees, Aur{\'e}lien and Xiao, Huangyu and Xue, Xiao",
    title = "{Probing Axions via Spectroscopic Measurements of S-stars at the Galactic Center}",
    eprint = "2507.07482",
    archivePrefix = "arXiv",
    primaryClass = "hep-ph",
    reportNumber = "FERMILAB-PUB-25-0150-T",
    month = "7",
    year = "2025"
}

@article{Li:2026dls,
    author = "Li, Ximeng and Zhong, Zhen and Chen, Yifan and Zhao, Yue and Omiya, Hidetoshi and Cardoso, Vitor",
    title = "{Dynamics and Frequency Conversion of Accreting Axion Clouds}",
    eprint = "2609.00128",
    archivePrefix = "arXiv",
    primaryClass = "hep-ph",
    month = "8",
    year = "2026"
}

@article{Chen:2019fsq,
    author = "Chen, Yifan and Shu, Jing and Xue, Xiao and Yuan, Qiang and Zhao, Yue",
    title = "{Probing Axions with Event Horizon Telescope Polarimetric Measurements}",
    eprint = "1905.02213",
    archivePrefix = "arXiv",
    primaryClass = "hep-ph",
    doi = "10.1103/PhysRevLett.124.061102",
    journal = "Phys. Rev. Lett.",
    volume = "124",
    number = "6",
    pages = "061102",
    year = "2020"
}

@article{Chen:2021lvo,
    author = "Chen, Yifan and Liu, Yuxin and Lu, Ru-Sen and Mizuno, Yosuke and Shu, Jing and Xue, Xiao and Yuan, Qiang and Zhao, Yue",
    title = "{Stringent axion constraints with Event Horizon Telescope polarimetric measurements of M87$^{*}$}",
    eprint = "2105.04572",
    archivePrefix = "arXiv",
    primaryClass = "hep-ph",
    doi = "10.1038/s41550-022-01620-3",
    journal = "Nature Astron.",
    volume = "6",
    number = "5",
    pages = "592--598",
    year = "2022"
}

@article{Chen:2022kzv,
    author = "Chen, Yifan and Xue, Xiao and Brito, Richard and Cardoso, Vitor",
    title = "{Photon Ring Astrometry for Superradiant Clouds}",
    eprint = "2211.03794",
    archivePrefix = "arXiv",
    primaryClass = "gr-qc",
    reportNumber = "DESY 22-170",
    doi = "10.1103/PhysRevLett.130.111401",
    journal = "Phys. Rev. Lett.",
    volume = "130",
    number = "11",
    pages = "111401",
    year = "2023"
}

@book{Maggiore:2007nxb,
    author = {Maggiore, Michele},
    title = {Gravitational Waves: Volume 1: Theory and Experiments},
    publisher = {Oxford University Press},
    year = {2007},
    month = {10},
    isbn = {9780198570745},
    doi = {10.1093/acprof:oso/9780198570745.001.0001},
    url = {https://doi.org/10.1093/acprof:oso/9780198570745.001.0001},
}

@article{Dupuis:2018dhs,
    author = "Dupuis, {\'E}ric and Paranjape, M. B.",
    title = "{New sources of gravitational wave signals: The black hole graviton laser}",
    eprint = "1807.03163",
    archivePrefix = "arXiv",
    primaryClass = "gr-qc",
    reportNumber = "UdeM-GPP-TH-18-262, UDEM-GPP-TH-18-262",
    doi = "10.1142/S0218271818470090",
    journal = "Int. J. Mod. Phys. D",
    volume = "27",
    number = "14",
    pages = "1847009",
    year = "2018"
}

@article{An:2026ugw,
    author = "An, Yu and Ge, Xian-Hui and Gong, Yun-Gui and Zhang, Yun-Long",
    title = "{Stimulated Emission from Boson Clouds}",
    eprint = "2606.08220",
    archivePrefix = "arXiv",
    primaryClass = "gr-qc",
    month = "6",
    year = "2026"
}

@article{Roy:2025qaa,
    author = "Roy, Soumen and Vicente, Rodrigo and Aurrekoetxea, Josu C. and Clough, Katy and Ferreira, Pedro G.",
    title = "{Scalar Fields around Black Hole Binaries in LIGO-Virgo-KAGRA}",
    eprint = "2510.17967",
    archivePrefix = "arXiv",
    primaryClass = "gr-qc",
    reportNumber = "MIT-CTP/5937",
    doi = "10.1103/fv9z-zkxx",
    journal = "Phys. Rev. Lett.",
    volume = "136",
    number = "19",
    pages = "191402",
    year = "2026"
}

@article{Liu:2021llm,
    author = "Liu, Tao and Lyu, Kun-Feng",
    title = "{The BH-PSR Gravitational Molecule}",
    eprint = "2107.09971",
    archivePrefix = "arXiv",
    primaryClass = "astro-ph.HE",
    month = "7",
    year = "2021"
}

@article{Dyson:2025dlj,
    author = "Dyson, Conor and Spieksma, Thomas F. M. and Brito, Richard and van de Meent, Maarten and Dolan, Sam",
    title = "{Environmental Effects in Extreme-Mass-Ratio Inspirals: Perturbations to the Environment in Kerr Spacetimes}",
    eprint = "2501.09806",
    archivePrefix = "arXiv",
    primaryClass = "gr-qc",
    doi = "10.1103/PhysRevLett.134.211403",
    journal = "Phys. Rev. Lett.",
    volume = "134",
    number = "21",
    pages = "211403",
    year = "2025"
}

@article{Li:2025ffh,
    author = "Li, Dongjun and Weller, Colin and Bourg, Patrick and LaHaye, Michael and Yunes, Nicol{\'a}s and Yang, Huan",
    title = "{Extreme mass-ratio inspiral within an ultralight scalar cloud: Scalar radiation}",
    eprint = "2507.02045",
    archivePrefix = "arXiv",
    primaryClass = "gr-qc",
    doi = "10.1103/7l9s-g21j",
    journal = "Phys. Rev. D",
    volume = "112",
    number = "8",
    pages = "084057",
    year = "2025"
}

@article{Tomaselli:2025zdo,
    author = "Tomaselli, Giovanni Maria and Caputo, Andrea",
    title = "{Probing dense environments around Sgr A* with S-star dynamics}",
    eprint = "2509.03568",
    archivePrefix = "arXiv",
    primaryClass = "astro-ph.GA",
    doi = "10.1103/pm7p-c53w",
    journal = "Phys. Rev. D",
    volume = "113",
    number = "8",
    pages = "083035",
    year = "2026"
}

@article{Tomaselli:2025jfo,
    author = "Tomaselli, Giovanni Maria",
    title = "{Smooth binary evolution from wide resonances in boson clouds}",
    eprint = "2507.15110",
    archivePrefix = "arXiv",
    primaryClass = "gr-qc",
    doi = "10.1103/h3fy-fyrx",
    journal = "Phys. Rev. D",
    volume = "112",
    number = "6",
    pages = "063033",
    year = "2025"
}

@article{Xu:2026cky,
    author = "Xu, Qi-Xuan and Brito, Richard and Della Monica, Riccardo and Vicente, Rodrigo and Yuan, Chen",
    title = "{Resonances as signatures of scalar clouds in eccentric extreme-mass-ratio inspirals}",
    eprint = "2605.03756",
    archivePrefix = "arXiv",
    primaryClass = "gr-qc",
    month = "5",
    year = "2026"
}

@article{Xu:2026aic,
    author = "Xu, Qi-Xuan and Brito, Richard and Della Monica, Riccardo and Vicente, Rodrigo and Yuan, Chen",
    title = "{Relativistic effects in extreme-mass-ratio inspirals within scalar clouds: Eccentric and inclined orbits}",
    eprint = "2606.21439",
    archivePrefix = "arXiv",
    primaryClass = "gr-qc",
    month = "6",
    year = "2026"
}

@article{DellaMonica:2025zby,
    author = "Della Monica, Riccardo and Brito, Richard",
    title = "{Detectability of gravitational atoms in black hole binaries with the Einstein Telescope}",
    eprint = "2503.23419",
    archivePrefix = "arXiv",
    primaryClass = "gr-qc",
    reportNumber = "ET-0056A-25",
    doi = "10.1103/h7ld-vv9p",
    journal = "Phys. Rev. D",
    volume = "112",
    number = "2",
    pages = "024074",
    year = "2025"
}

@article{Kim:2025wwj,
    author = "Kim, Hyungjin and Lenoci, Alessandro",
    title = "{Self-gravity in superradiance clouds: Implications for binary dynamics and observational prospects}",
    eprint = "2508.08367",
    archivePrefix = "arXiv",
    primaryClass = "gr-qc",
    reportNumber = "DESY-25-116",
    doi = "10.1103/81dj-kxmy",
    journal = "Phys. Rev. D",
    volume = "112",
    number = "10",
    pages = "104014",
    year = "2025"
}

@article{PhysRevLett.40.223,
  title = {A New Light Boson?},
  author = {Weinberg, Steven},
  journal = {Phys. Rev. Lett.},
  volume = {40},
  issue = {4},
  pages = {223--226},
  numpages = {0},
  year = {1978},
  month = {Jan},
  publisher = {American Physical Society},
  doi = {10.1103/PhysRevLett.40.223},
  url = {https://link.aps.org/doi/10.1103/PhysRevLett.40.223}
}

@article{Ferreira:2020fam,
    author = "Ferreira, Elisa G. M.",
    title = "{Ultra-light dark matter}",
    eprint = "2005.03254",
    archivePrefix = "arXiv",
    primaryClass = "astro-ph.CO",
    doi = "10.1007/s00159-021-00135-6",
    journal = "Astron. Astrophys. Rev.",
    volume = "29",
    number = "1",
    pages = "7",
    year = "2021"
}

@article{Hui:2016ltb,
    author = "Hui, Lam and Ostriker, Jeremiah P. and Tremaine, Scott and Witten, Edward",
    title = "{Ultralight scalars as cosmological dark matter}",
    eprint = "1610.08297",
    archivePrefix = "arXiv",
    primaryClass = "astro-ph.CO",
    doi = "10.1103/PhysRevD.95.043541",
    journal = "Phys. Rev. D",
    volume = "95",
    number = "4",
    pages = "043541",
    year = "2017"
}

@article{Marsh:2015xka,
    author = "Marsh, David J. E.",
    title = "{Axion Cosmology}",
    eprint = "1510.07633",
    archivePrefix = "arXiv",
    primaryClass = "astro-ph.CO",
    reportNumber = "KCL-PH-TH-2015-50",
    doi = "10.1016/j.physrep.2016.06.005",
    journal = "Phys. Rept.",
    volume = "643",
    pages = "1--79",
    year = "2016"
}

\end{document}